\documentclass{imamat}

\jno{drnxxx}              
\received{XX XX XXXX}     
\revised{XX XX XXXX}      
\accepted{XX XX XXXX}     

\usepackage{amsmath} 
\usepackage{amssymb}
\usepackage{amsthm} 
\usepackage{amsfonts}
\usepackage{graphicx} 

\usepackage{color}

\usepackage{tikz}
\usepackage{pgfplots}
\usepackage{pgfplotstable}
\usetikzlibrary{calc}
\usetikzlibrary{arrows}
\usetikzlibrary{spy}
\usetikzlibrary{decorations.markings}
\usepgfplotslibrary{groupplots}

\pgfrealjobname{paper} 

\pgfplotsset{compat=1.17} 

\tikzset{->-/.style={decoration={
  markings,
  mark=at position #1 with {\arrow{>}}},postaction={decorate}}}
\tikzset{-<-/.style={decoration={
  markings,
  mark=at position #1 with {\arrow{<}}},postaction={decorate}}}  

\DeclareMathOperator{\sech}{sech}

\begin{document}

\title{Localised patterns in a generalised Swift--Hohenberg equation with a 
       quartic marginal stability curve$^\dag$}
\shorttitle{Localised patterns with a quartic marginal stability curve} 
\shortauthorlist{Bentley \& Rucklidge} 

\author{%
\name{David C. Bentley}
\and
\name{Alastair M. Rucklidge$^*$}
\address{School of Mathematics, University of Leeds, Leeds LS2 9JT, 
         UK\email{$^*$Corresponding author: {\texttt{A.M.Rucklidge@leeds.ac.uk}}}}}

\maketitle

\begin{abstract}
{In some pattern-forming systems, for some parameter values, patterns form with
two wavelengths, while for other parameter values, there is only one
wavelength. The transition between these can be organised by a
codimension-three point at which the marginal stability curve has a quartic
minimum. We develop a model equation to explore this situation, based on the
Swift--Hohenberg equation; the model contains, amongst other things, snaking
branches of patterns of one wavelength localised in a background of patterns of
another wavelength. In the small-amplitude limit, the amplitude equation for
the model is a generalised Ginzburg--Landau equation with fourth-order spatial
derivatives, which can take the form of a complex Swift--Hohenberg equation
with real coefficients. Localised solutions in this amplitude equation help
interpret the localised patterns in the model. This work extends recent efforts 
to investigate snaking behaviour in pattern-forming systems where two different 
stable non-trivial patterns exist at the same parameter values.}
 {Pattern formation, quartic minimum, two wavelengths, localised patterns}
\\
2020 Math Subject Classification: 35B36, 35B32, 37G05, 37L10 \\
\\
$\dag$~This paper is dedicated to the memory of Thomas Wagenknecht (1974--2012).
\end{abstract}



\section{Introduction}
\label{sec:introduction}

Pattern formation most commonly occurs with a single wavelength, as in for
example zebra stripes, Rayleigh--B\'{e}nard convection and the Taylor--Couette
flow \citep{Hoyle2006}. In these examples, there is a featureless basic state
that loses stability to waves with a nonzero wavelength as a control parameter
is increased. Typically the marginal stability curve, which separates stable
from unstable waves depending on their wavelength and the control parameter,
has a quadratic minimum.

In recent years, it has been recognised that pattern formation with two length
scales can lead to a wide variety of complex and interesting patterns, such as
superlattice patterns, quasipatterns and spatio-temporal chaos \cite[see for 
example][and
references therein]{Castelino2020}. Having two length scales can arise in
different ways: in the Faraday wave problem with multi-frequency forcing, for
example, patterns with the two length scales arise in response to different
components of the forcing
\citep{Edwards1994,Topaz2002,Rucklidge2009,Skeldon2015}. Another possibility is
that the quadratic minimum in the marginal stability curve can change to a
quadratic maximum with two nearby quadratic minima at the two length scales.
This transition can occur via a quartic minimum, and is found in the
magnetised Taylor--Couette experiement \citep{Stefani2009,Mamatsashvili2019},
Lapwood--Prats convection \citep{Rees2013}, binary phase field crystals
\citep{Holl2020}, surface waves in ferrofluids \citep{Raitt1997} and nonlinear
optics \citep{Kozyreff2009}. This paper is concerned first with developing and
analysing a model that contains this transition in as simple a form as
possible, and second with investigating localised patterns in the model.

Problems with a single length scale where the pattern-forming bifurcation is
subcritical can have parameter intervals where both the featureless and the
patterned solutions are stable. In this case, it is possible to find localised
solutions consisting of a region of a spatially periodic pattern embedded in a
spatially homogeneous background
\cite[see][for reviews]{Dawes2010,Knobloch2015}. With two length scales,
there is a wider variety of possibilities, including having patterns with one
wavelength embedded in a pattern with a different wavelength. This phenomenon
has been observed in Rayleigh--B\'{e}nard convection in a long, thin channel,
or slot \citep{Hegseth1992} and has been explored in the context of 
generalised Ginzburg--Landau models
\citep{Riecke1990,Raitt1995,Raitt1997,Kozyreff2009}.

A useful mathematical tool for studying pattern formation is the construction
of model equations that display qualitatively similar behaviour as the physical
system under consideration, but whose analysis is more tractable. Perhaps the
most ubiquitous of such model equations is the Swift--Hohenberg (SH) equation
\citep{Swift1977}, originally introduced as a model of thermal
fluctuations near the onset of Rayleigh--B\'{e}nard convection. It has been
used extensively in the study of localised patterns, starting with the work of
\citet{Hilali1995}, \citet{Crawford1999b} and \citet{Woods1999}. The equation
(in one dimension) is
 \begin{equation}
 \label{eq:SHE}
 \frac{\partial u}{\partial t} = \mu u - 
      \left( 1 + \frac{\partial^{2}}{\partial x^{2}} \right)^{2} u + 
      n_{2} u^{2} + n_{3} u^{3}, 
 \end{equation}
where $u(x,t)\in\mathbb{R}$ represents the pattern, $\mu$~is the driving
parameter, and $n_2$ and $n_3$ are parameters controlling the nonlinear terms
(typically $n_3=-1$). We consider equation~\eqref{eq:SHE} subject to periodic
boundary conditions on a domain $x\in[0,L]$.

The featureless (or trivial) solution $u=0$ is stable for~$\mu<0$. Small
amplitude perturbations of the form $e^{\sigma t+ikx}$ grow as
$\sigma=\mu-\left(1-k^{2}\right)^{2}$, so for $\mu>0$ the maximum growth rate is
at critical wavenumber $k=1$ (independent of~$\mu$), and if $\mu>0$, a range of
wavenumbers will grow exponentially until nonlinear effects become important.
The marginal stability curve is found by setting $\sigma=0$, so
 \begin{equation*}
 \mu = \left( 1 - k^{2} \right)^{2},
 \end{equation*}
which has a quadratic minimum at $k=1$.
At $\mu=0$ the trivial solution undergoes a pitchfork bifurcation, creating a
branch of spatially periodic solutions, which is stable if it is 
supercritical, and unstable if not.

In large domains ($L\gg1$), and with small amplitude solutions ($u={\mathcal 
O}(\epsilon)$), standard weakly nonlinear theory can be applied. The pattern
is written, with scaled space, time and parameter, as
 \begin{equation}
 \label{eq:GLE_scaling}
 u(x,t) = \epsilon A(X,T) e^{ix} + \text{c.c.} + \text{h.o.t.},
 \quad
 X = \epsilon x,
 \quad
 T = \epsilon^2 t,
 \quad
 \mu = \epsilon^2 \mu_2,
 \end{equation}
where $\text{c.c.}$ refers to the complex conjugate and $\text{h.o.t.}$
refers to higher-order terms. In this limit, the solvability condition for~$A$ at 
third-order in~$\epsilon$ results in the Ginzburg--Landau (GL) equation
\citep{Cross1993}:
 \begin{equation}
 \label{eq:GLE}
 A_T = \mu_2 A + 4 A_{XX} + \left(3 n_3 + \tfrac{38}{9}n_2^2\right) |A|^2A,
 \end{equation}
where subscripts $T$ and $X$ refer to partial derivatives.
This equation governs the long-wavelength slow evolution of
the amplitude of solutions of~\eqref{eq:SHE}.
We define the coefficient of the nonlinear term
 \begin{equation*}
 n_{SH} = 3 n_{3} + \tfrac{38}{9} n_{2}^{2},
 \end{equation*}
which determines the criticality of the bifurcation at $\mu = 0$: if $n_{SH}<0$
the bifurcation is supercritical, and if $n_{SH}>0$, the bifurcation is
subcritical. In the supercritical case, the GL equation gives nonlinear
stability of striped patterns to long-wavelength perturbations \citep[the
Eckhaus instability, see][]{Eckhaus1965}. In the subcritical case, the GL
equation allows localised $\sech$-profile solutions to
equation~\eqref{eq:SHE} that can be continued in~$\mu$, leading to the
well known homoclinic ``snaking'' structure of localised solutions of the
Swift--Hohenberg and many other pattern forming
problems~\citep{Woods1999,Burke2006,Burke2007,Beck2009,Chapman2009}.

The GL equation~\eqref{eq:GLE} is one standard tool useful in the analysis of
the SH equation~\eqref{eq:SHE}. There are three others that we mention briefly. 
First, the SH equation as written above is variational in time, and admits 
a Lyapunov functional:
 \begin{equation}
 \label{eq:SH_Lyaponov}
 \mathcal{F}[u]
      = \int_{0}^{L} \left( - \tfrac{1}{2} \mu u^{2} 
                            + \tfrac{1}{2} \left( \left( 1 + \partial_{x}^{2} \right) u \right)^{2} 
                            - \tfrac{1}{3}n_{2} u^{3} 
                            - \tfrac{1}{4}n_{3} u^{4} \right) dx.
 \end{equation}
With $n_3<0$, it can easily be shown that $\mathcal{F}[u]$ is bounded below and 
that it is a decreasing function of time:
 \begin{equation*}
 \frac{d \mathcal{F}}{d t} = 
      - \int_{0}^{L} \left( \frac{\partial u}{\partial t} \right)^{2} dx
      \leq 0.
 \end{equation*}
Equilibrium states correspond to stationary points of~$\mathcal{F}$, and those
coinciding with local minima of $\mathcal{F}$ must necessarily be stable. A
front connecting two patterns with different values of~$\mathcal{F}$ will tend
to move towards (and so eliminate) the pattern with the larger value. Localised
solutions are found near the Maxwell point, where the pattern and the zero
state have the same value of~$\mathcal F$, and near this point the difference
in~$\mathcal{F}$ is small enough that the front becomes pinned
\citep{Pomeau1986} to the underlying pattern.

The second tool is the observation that the steady 
Swift--Hohenberg equation~\eqref{eq:SHE} admits a first integral in space. 
Multiplying the time-independent version of~\eqref{eq:SHE}
through by $-u_{x}$, and integrating with respect to~$x$, yields
 \begin{align*}
 \mathcal{H} &= - \int \left( \mu u - u - 2 u_{xx} - u_{xxxx} + n_{2} u^{2} + n_{3} u^{3} \right) u_{x} dx \nonumber \\
 &= - \tfrac{1}{2} \left( \mu - 1 \right) u^{2} + u_{x}^{2} 
    - \tfrac{1}{2} u_{xx}^{2} + u_{x} u_{xxx} 
    - \tfrac{1}{3}n_{2} u^{3} - \tfrac{1}{4}n_{3} u^{4},
 \end{align*}
 and so $\frac{d\mathcal{H}}{dx}=0$. The quantity $\mathcal{H}$ is sometimes
referred to as the Hamiltonian for the steady version of~\eqref{eq:SHE}, since
there is a change of coordinates under which the system has Hamiltonian
structure. If there is a steady front connecting two patterns, the condition 
$\frac{d\mathcal{H}}{dx}=0$ means that the two patterns must have the same value
of~$\mathcal{H}$.

The third useful tool is to note, again for the time-independent version
of~\eqref{eq:SHE}, that there is a Hamiltonian--Hopf bifurcation in space as
$\mu$~crosses zero. At the bifurcation point, there is a pair of double spatial
eigenvalues~$\pm{i}$, and the normal form can be written as a pair of
first-order ODEs in~$x$ for two complex variables. A bifurcation analysis
performed by \citet{Iooss1993} and \citet{Iooss1998a}, and extended by
\citet{Woods1999}, provides a geometrical interpretation of the solutions
of the normal form. In particular, there are parameter values where there are
solutions of the SH equation that are homoclinic to the origin as
$x\to\pm\infty$ \citep{Burke2007a}: these homoclinic orbits represent
localised solutions.

The existence and bifurcation structure of localised solutions in the
Swift--Hohenberg equation is now well understood: see 
\citet{Dawes2010} and \cite{Knobloch2015} for reviews. In this paper, we
consider a generalised version of the Swift--Hohenberg equation that allows a
quartic minimum of the marginal stability curve ($\S$\ref{sec:model_eqn}).
Unfolding this quartic minimum, and using tools such as generalised versions of
the Ginzburg--Landau equation ($\S$\ref{sec:wnlt}), the Lyapunov function and
the first integral introduced above ($\S$\ref{sec:Lyapunov_first_int}), allows
us to identify parameter regimes where we can find patterns of one wavenumber
localised in a background of patterns with a different wavenumber
($\S$\ref{sec:localised}). We discuss the significance of our results in
$\S$\ref{sec:conclusion}. We include normal form calculations in
Appendix~\ref{sec:appendix}, but we have not found first integrals of the
normal form, and so we have not been able to put it to immediate use.
                               

\section{The model equation}
\label{sec:model_eqn}

In this section we build a model equation to explore the unfolding of the
quartic minimum in the marginal stability curve. We start with the 
SH~equation~\eqref{eq:SHE}, modified to allow a marginal 
stability that can change from having one to having two minima, and then
add a selection of nonlinear terms.

\subsection{Linear terms}
\label{subsec:linear_terms}

As a starting 
point, we consider a linear part of the PDE based on the polynomial 
 \begin{equation*}
 p_0(K) = \left( 1 - K \right)^{4},
 \end{equation*}
where $K=k^{2}$, so $1-K$ will become $1 + \partial_{xx}$ in the model equation. 
The quartic minimum at $K=1$ can be unfolded by adding two small terms to the
equation, yielding
 \begin{equation}
 \label{eq:polynomial}
 p(K) = p_0(K) + f_1 K + f_2 K^2,
 \end{equation}
with $f_1=f_2=0$ at the quartic minimum.
In principle, small terms $f_0$, $f_3K^3$ and $f_4K^4$ could be added as well,
but $f_0$~can be absorbed into the bifurcation parameter~$\mu$, $f_3$~can be
eliminated by making a small shift in $K$ by~$-\frac{1}{4}f_3$, and $f_4$ can be 
absorbed by an overall scaling.

Before writing down the PDE model, we consider the transition from having one
minimum to two. The condition that $p(K)$ has two minima with a maximum in
between is the same as the condition that the derivative $p'(K)$ has three 
distinct real
roots. Now
 \begin{equation*}
 p'(K) = 4K^3 - 12K^2 + (2f_2+12)K + f_1-4,
 \end{equation*}
and the condition that a cubic polynomial has three distinct real roots is that 
its discriminant should be positive. The boundary, where the discriminant is 
zero, occurs where
 \begin{equation*}
 f_1^2  + 4f_1f_2 + 4f_2^2 + \tfrac{8}{27}f_2^3  = 0,
 \quad\text{or}\quad
 f_1 = -2f_2 \pm \left(-\tfrac{2}{3}f_2\right)^{\frac{3}{2}}.
 \end{equation*}                               
In addition, when the discriminant is positive, $p(K)$~has two minima at $K_1$ and $K_3$ 
with a maximum at $K_2$ in between, with $K_1<K_2<K_3$, $K_1+K_2+K_3=3$ and
$K_1K_2K_3=1-\frac{1}{4}f_1$ (found 
from the relationship between the roots and coefficients of the cubic $p'(K)=0$).
Manipulating the conditions $p'(K_1)=0$, $p'(K_3)=0$ and $p(K_1)=p(K_3)$ leads 
to the conclusion that the two minima are equal when
 \begin{equation*}
 K_1 + K_3 = 2,\quad
 K_2 = 1, \quad
 \text{and}\quad
 f_1 + 2f_2 = 0,
 \end{equation*}
that is, the intermediate maximum is at $K=1$ and the two minima are equally
spaced on each side.
 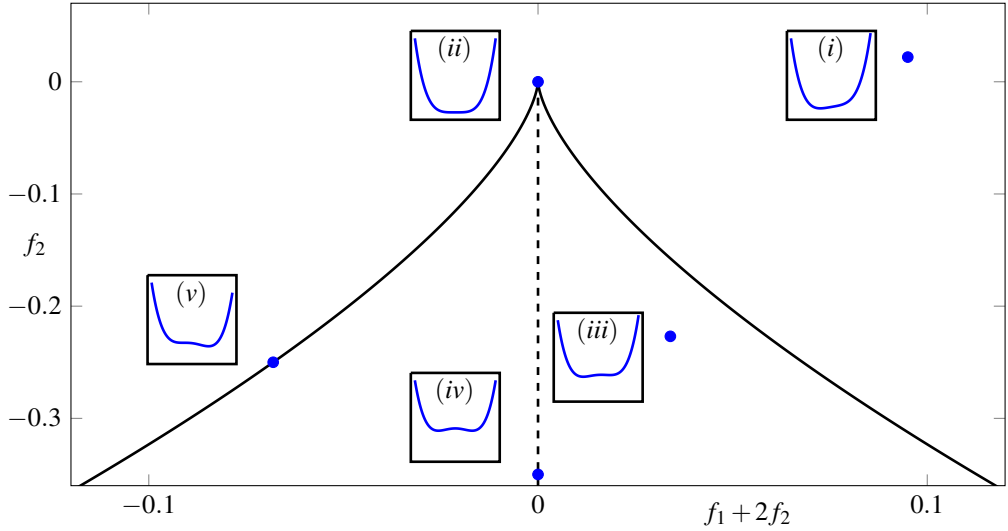
\begin{figure}
 \begin{center}
 \mbox{\beginpgfgraphicnamed{paper_figure_01}%
  \begin{tikzpicture}
  \begin{axis}[width=14cm,height=8cm,xmin=-0.12,xmax=0.12,ymin=-0.36,ymax=0.07,xtick={-0.1,0,0.1}]
  \addplot[color=black, mark=none, line width=1pt] 
          table [x index=4,y index=0] {paper-cusp.dat};
  \addplot[color=black, mark=none, line width=1pt] 
          table [x index=5,y index=0] {paper-cusp.dat};
  \addplot[color=black, mark=none, line width=1=0.5pt, dashed] 
          table [x index=6,y index=0] {paper-cusp.dat};
%
%
  \addplot[color=blue,mark=*] coordinates{(0.095,0.022)};       
  \addplot[color=blue,mark=*] coordinates{(0.00, 0.00)};        
  \addplot[color=blue,mark=*] coordinates{(0.0340,-0.2269)};    
  \addplot[color=blue,mark=*] coordinates{(0.000,-0.35)};       
  \addplot[color=blue,mark=*] coordinates{(-0.0680414,-0.25)};  
 \end{axis}
\draw (10.1,5.5) node[above] {$(i)$};
\draw (5.1,5.5) node[above] {$(ii)$};
\draw (7.0,1.75) node[above] {$(iii)$};
\draw (5.1,0.95) node[above] {$(iv)$};
\draw (1.6,2.25) node[above] {$(v)$};
\draw (9.0,-0.1) node[below] {$f_{1}+2f_{2}$};
\draw (-0.2,3.2) node[left] {$f_{2}$};
\begin{axis}[axis lines=none,xtick=\empty,ytick=\empty,scaled ticks=false,xshift=9.4cm,yshift=4.75cm,width=3cm,height=3cm]
 \addplot[mark=none,samples=20,color=blue,line width=1pt] table[x index=0,y index=1] {paper-cusp_curve_i.dat};
 \addplot[mark=none,color=black,line width=1=0.5pt] coordinates{(-0.1,1.1) (-0.1,-0.1) (2.1,-0.1) (2.1,1.1) (-0.1,1.1)};
\end{axis}
\begin{axis}[axis lines=none,xtick=\empty,ytick=\empty,scaled ticks=false,xshift=4.4cm,yshift=4.75cm,width=3cm,height=3cm]
 \addplot[mark=none,samples=20,color=blue,line width=1pt] table[x index=0,y index=1] {paper-cusp_curve_ii.dat};
 \addplot[mark=none,color=black,line width=1=0.5pt] coordinates{(-0.1,1.1) (-0.1,-0.1) (2.1,-0.1) (2.1,1.1) (-0.1,1.1)};
\end{axis}
\begin{axis}[axis lines=none,xtick=\empty,ytick=\empty,scaled ticks=false,xshift=6.3cm,yshift=1cm,width=3cm,height=3cm]
 \addplot[mark=none,samples=20,color=blue,line width=1pt] table[x index=0,y index=1] {paper-cusp_curve_iii.dat};
 \addplot[mark=none,color=black,line width=1=0.5pt] coordinates{(-0.1,1.1) (-0.1,-0.1) (2.1,-0.1) (2.1,1.1) (-0.1,1.1)};
\end{axis}
\begin{axis}[axis lines=none,xtick=\empty,ytick=\empty,scaled ticks=false,xshift=4.4cm,yshift=0.2cm,width=3cm,height=3cm]
 \addplot[mark=none,samples=20,color=blue,line width=1pt] table[x index=0,y index=1] {paper-cusp_curve_iv.dat};
 \addplot[mark=none,color=black,line width=1=0.5pt] coordinates{(-0.1,1.1) (-0.1,-0.1) (2.1,-0.1) (2.1,1.1) (-0.1,1.1)};
\end{axis}
\begin{axis}[axis lines=none,xtick=\empty,ytick=\empty,scaled ticks=false,xshift=0.9cm,yshift=1.5cm,width=3cm,height=3cm]
 \addplot[mark=none,samples=20,color=blue,line width=1pt] table[x index=0,y index=1] {paper-cusp_curve_v.dat};
 \addplot[mark=none,color=black,line width=1=0.5pt] coordinates{(-0.1,1.1) (-0.1,-0.1) (2.1,-0.1) (2.1,1.1) (-0.1,1.1)};
\end{axis}
   \end{tikzpicture}
 \endpgfgraphicnamed}
   \end{center}
 \vspace{-3ex}
 \caption{\label{fig:cusp} Five examples of the marginal stability
 curves $\mu=p(K)$ in different regions of the 
 $(f_{1}+2f_{2},f_{2})$ parameter space. 
 The curves (blue in black boxes) shown are:
 $(i)$~single minimum, 
 $(ii)$~quartic minimum, 
 $(iii)$~double minima, with the left minimum being the lower, 
 $(iv)$~double minima, with both minima at the same height, and 
 $(v)$~a single minimum and an inflexion point. The specific parameters for each 
example are shown as blue dots.
The solid lines indicate where 
the discriminant is zero (the derivative has a double zero), and the dashed 
line ($f_{1}+2f_{2}=0$, $f_2<0$) indicates where the two minima exist and have the same height.}
\end{figure}

The next step is to convert the polynomial to the linear operator of the model
equation:
 \begin{equation*}
 \frac{\partial u}{\partial t} = \mu u - \left( 1 + \frac{\partial^{2}}{\partial x^{2}} \right)^{4} u + f_{1} \frac{\partial^{2} u}{\partial x^{2}} - f_{2} \frac{\partial^{4} u}{\partial x^{4}}.
 \end{equation*}
A mode $u=e^{\sigma t+ikx}$ has growth rate
$\sigma=\mu-(1-k^2)^4-f_1k^2-f_2k^4=\mu-p(k^2)$, connecting the dispersion
relation to the polynomial~$p(K)$ discussed above. Marginal
stability, when $\sigma=0$, occurs when $\mu=p(k^2)$, and Figure~\ref{fig:cusp}
shows examples of the marginal stability curves in the
$(f_1+2f_2,f_2)$ parameter plane. The discriminant is positive, and there is a
double minimum, within the cusp-shaped region below the solid curves, and the
two minima are equal on the dashed line.
The cusp, the point at which there is a quartic minimum with $f_{1}=f_{2}=0$,
represents a codimension-three bifurcation as $\mu$~crosses zero.

\begin{figure}
 \begin{center}
 \mbox{\beginpgfgraphicnamed{paper_figure_02}%
  \begin{tikzpicture}
   \begin{axis}[width=6.5cm,height=5cm,
                scaled ticks=false,
                xmin=0.8,xmax=1.2,
                xtick={0.8,1,1.2},
                ymin=-0.0001,ymax=0.002,
                ytick={0,0.001,0.002},
                yticklabel style={anchor=east,
                                  /pgf/number format/precision=3,
                                  /pgf/number format/fixed,
                                  /pgf/number format/fixed zerofill},
               ]
    \addplot[color=blue, mark=none, line width=1pt] 
             table [x index=0,y index=1] {paper-proctor.dat};
    \addplot[color=blue, mark=none, line width=1pt, dashed] 
             table [x index=0,y index=2] {paper-proctor.dat};
   \end{axis}
\draw (4.0, 0.0) node[below] {$k^2$};
\draw (0, 2.75) node[left] {$\mu$};
\draw (0.25, 3.5) node[above] {(a)};
  \end{tikzpicture}
\hskip 0.5truecm
  \begin{tikzpicture}
   \begin{axis}[width=6.5cm,height=5cm,
                scaled ticks=false,
                xmin=0.8,xmax=1.2,
                xtick={0.8,1,1.2},
                ymin=0.0259,ymax=0.028,
                ytick={0.026,0.027,0.028},
                yticklabel style={anchor=east,
                                  /pgf/number format/precision=3,
                                  /pgf/number format/fixed,
                                  /pgf/number format/fixed zerofill},
               ]
    \addplot[color=blue, mark=none, line width=1pt] 
             table [x index=0,y index=3] {paper-proctor.dat};
    \addplot[color=blue, mark=none, line width=1pt, dashed] 
             table [x index=0,y index=4] {paper-proctor.dat};
   \end{axis}
\draw (4.0, 0.0) node[below] {$k^2$};
\draw (0, 2.75) node[left] {$\mu$};
\draw (0.25, 3.5) node[above] {(b)};
  \end{tikzpicture}
 \endpgfgraphicnamed}
 \end{center}
\caption{Marginal stability curve (solid) and curve of maximum (or minimum) 
 growth (dashed) for~\eqref{eq:PDE_linear_Proctor}, for 
 (a)~$f_{1} = f_{2} = 0$, and
 (b)~$f_{1}=0.05$ and $f_{2}=-0.0235$.
 The coefficient $\mu_{p} = -0.1$ in both cases.
 Two of the three dashed lines in~(b) join together above the top of the frame.
 \label{fig:Proctor_curves}}
\end{figure}
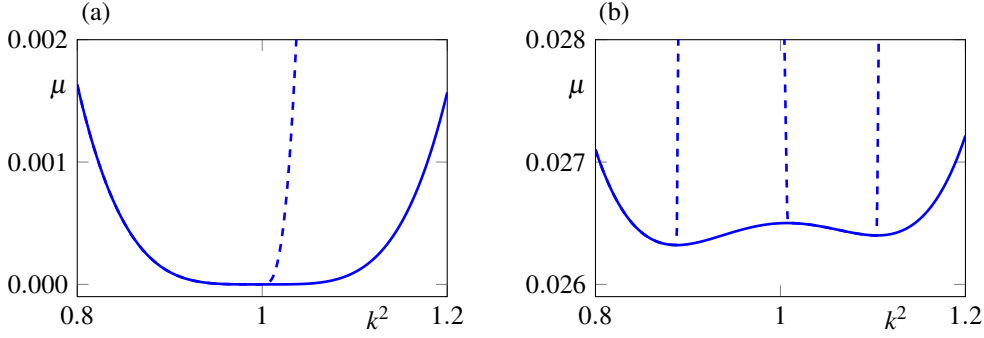

However, a model based solely on this would have a wavenumber for maximum growth
rate for solutions that did not depend on~$\mu$. 
Regions of secondary instability of patterns 
are organised around the curve of maximum growth rate, and it is important that
this curve is modelled correctly.
In pattern-forming problems with a quadratic minimum, the
wavenumber for maximum growth rate typically depends only linearly on~$\mu$, so
with $\mu \ll 1$, this lack of dependence on~$\mu$ in the SH~equation is
reasonable. However, with a quartic minimum, \citet{Proctor1991} argues that
additional terms should be included in the unfolding of a quartic minimum in
order to allow the wavenumber of maximum growth rate to depend on~$\mu$. For
this reason,  we propose the modified linear operator
 \begin{equation}
 \label{eq:PDE_linear_Proctor}
 \frac{\partial u}{\partial t} = 
   \mu \left( 1 + \mu_{p} \left( 1 + \frac{\partial^{2} }{\partial x^{2}} \right) \right) u - \left( 1 + \frac{\partial^{2}}{\partial x^{2}} \right)^{4} u + f_{1} \frac{\partial^{2} u}{\partial x^{2}} - f_{2} \frac{\partial^{4} u}{\partial x^{4}},
 \end{equation}
where we will call the extra term, proportional to~$\mu_{p}$, the Proctor term,
although the form of this extra term differs from that proposed by 
\citet{Proctor1991}.
The dispersion relation is now
 \[
 \sigma = \mu \left( 1 + \mu_{p} \left( 1 - k^{2} \right) \right) - \left( 1 - k^{2} \right)^{4} - f_{1} k^{2} - f_{2} k^{4},
 \]
and so
 \[
 \frac{d \sigma}{d k^{2}} = - \mu \mu_{p} + 4 \left( 1 - k^{2} \right)^{3} - f_{1} - 2 f_{2} k^{2}.
 \]
The curve of maximum (or minimum) growth rate is then defined by 
$\frac{d\sigma}{dk^{2}}=0$, \textit{i.e.},
 \begin{equation*}
 \mu = \frac{1}{\mu_{p}} \left( 4 \left( 1 - k^{2} \right)^{3} - f_{1} - 2 f_{2} k^{2} \right).
 \end{equation*}
For $\mu_{p}\neq0$ and $f_1=f_2=0$ (Figure~\ref{fig:Proctor_curves}$a$), 
this curve of maximum growth rate is a cubic and so is tangent to the neutral 
stability curve, while for non-zero $f_1$ and $f_2$
(Figure~\ref{fig:Proctor_curves}$b$), the curves of maximum growth rate
intersect the two minima in the marginal stability curve transversally.

Including the Proctor term influences the shape of the marginal stability
curve. Throughout most of this paper we will assume that $\mu_{p}$ is small
enough not to influence the behaviour of solutions, apart from the asymptotic
analysis in~$\S$\ref{subsec:Proctor_term} (where the Proctor term necessarily
appears as a higher order term), and a numerical consideration
in~$\S$\ref{subsec:snaking}.

The linear part of the model derived so far is symmetric under spatial
reversals, \textit{i.e.}, $x$~and $-x$ are equivalent. Some systems break
reflection symmetry, for example, the Taylor--Couette system in the presence of
an azimuthal and axial magnetic field \citep{Stefani2009,Mamatsashvili2019};
this can be modelled by including terms such as $\left(f_{d0} -
f_{d1}\partial_{xx}\right)u_x$, leading to drifting solutions. However, the
reflection symmetry is useful in the analysis of the model equation, so we do
not include drift terms here.

\subsection{Nonlinear terms}
\label{subsec:nonlinear_terms}

Nonlinear terms in the model will saturate exponentially growing solutions at
finite amplitude, and should respect the symmetry (or lack thereof) of any
underlying physical systems. The usual SH~nonlinear term 
is~$-u^3$, allowing $u\rightarrow-u$ symmetry, but here we break this symmetry 
and chose 
 \[
 n_{2} u^{2} + n_{3} u^{3}.
 \]
as nonlinear terms for the model. There are many more possibilities for
nonlinear terms that have been used by many authors
\citep{Knobloch1990d,Crawford1999b,Burke2012} for similar equations in
different contexts. For the purposes of this paper, we consider
only the nonlinear terms above, retaining the coefficients $n_2$ and $n_3$ as
parameters. We focus on the case where the bifurcation is supercritical, which
limits the range of values that $n_{2}$ and $n_{3}$ can take. In particular, we
must have
 \begin{equation}
 \label{eq:nA}
 n_A \equiv 3 n_{3} + \tfrac{326}{81} n_{2}^{2} < 0, 
 \end{equation}
derived in section~\ref{subsec:derivation_GLE}. For numerical examples, we will 
take $n_{2} = 0.1$ and $n_{3} = -1$ throughout.

We now have our complete model equation, written here with the Proctor term
 \begin{equation}
 \label{eq:PDE_nonlinear_Proctor}
 \frac{\partial u}{\partial t} = 
   \mu \left( 1 + \mu_{p} \left( 1 + \frac{\partial^{2} }{\partial x^{2}} \right) \right) u - \left( 1 + \frac{\partial^{2}}{\partial x^{2}} \right)^{4} u + f_{1} \frac{\partial^{2} u}{\partial x^{2}} - f_{2} \frac{\partial^{4} u}{\partial x^{4}},
   + n_{2} u^{2} + n_{3} u^{3},
 \end{equation}
and without
 \begin{equation}
 \label{eq:PDE_nonlinear}
 \frac{\partial u}{\partial t} = \mu u - \left( 1 + \frac{\partial^{2}}{\partial x^{2}} \right)^{4} u + f_{1} \frac{\partial^{2} u}{\partial x^{2}} - f_{2} \frac{\partial^{4} u}{\partial x^{4}} + n_{2} u^{2} + n_{3} u^{3}.
 \end{equation}
For the remainder of this paper, we will concentrate mostly on the parameter
values where the marginal stability curves have two minima and there is
bistability between patterns of different wavelength, leading to the
possibility of these patterns coexisting in separate parts of the domain.
Although these equations are related to the two length scale models of 
\citet{Lifshitz1997} and \citet{Rucklidge2012}, our model cannot be derived 
from these simply by setting the two length scales to be equal.


\section{Weakly nonlinear analysis}
\label{sec:wnlt}

In this section we compute weakly nonlinear solutions for the
model~\eqref{eq:PDE_nonlinear} by deriving a generalised version of the
Ginzburg--Landau equation, and use it to establish where one-dimensional
periodic patterns are stable to long-wave perturbations. For most of the
calculations we set $\mu_{p}=0$, but we do consider the effect of the Proctor
term briefly in~$\S$\ref{subsec:Proctor_term}. We also compare our derivation
to previously published derivations for related problems
in~$\S$\ref{subsec:Kozyreff}.
 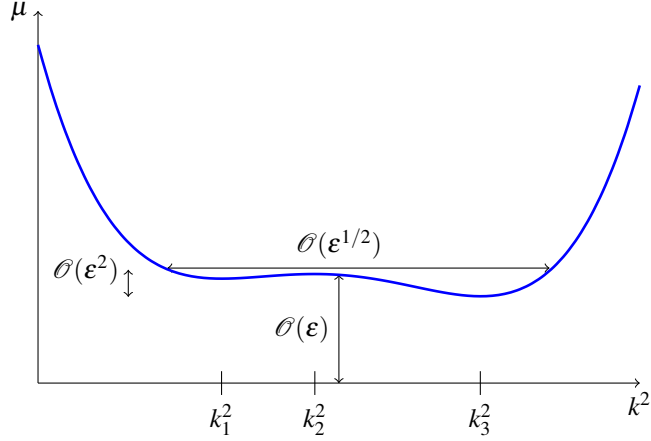
\begin{figure}
 \begin{center}
 \mbox{\beginpgfgraphicnamed{paper_figure_03}%
 \mbox{\begin{tikzpicture}[xscale=4.0,yscale=4.5]
  \draw[->] (0,0) -- (2,0) node[below] {$k^{2}$};
  \draw[->] (0,0) -- (0,1.1) node[left] {$\mu$};
  \draw[<->] (0.43,0.34) -- (1.7,0.34);
  \draw[<->] (0.3,0.255) -- (0.3,0.335) node[left] {$\mathcal{O}(\epsilon^{2})$};
  \draw[<->] (1,0) -- (1,0.32);
  \draw (1,0.16) node[left] {$\mathcal{O}(\epsilon)$};
  \draw[line width=1pt,color=blue] plot[id=pde_scales2, domain=0.0:2, samples=50, smooth]
          function{1 - 4*x + 6*x**2 - 4*x**3 + x**4 + 0.7*x - 0.38*x**2};
  \draw (0.61,-1pt) -- (0.61,1pt);
  \draw (0.92,-1pt) -- (0.92,1pt);
  \draw (1.47,-1pt) -- (1.47,1pt);
  \draw (0.61,-1pt) node[below] {$k_{1}^{2}$};
  \draw (0.92,-1pt) node[below] {$k_{2}^{2}$};
  \draw (1.47,-1pt) node[below] {$k_{3}^{2}$};
  \draw (1,0.34) node[above] {$\mathcal{O}(\epsilon^{1/2})$};
\end{tikzpicture}}
 \endpgfgraphicnamed}%
 \end{center}
 \caption{\label{fig:marginal_scalings} Scalings used for the asymptotic analysis.
 The deviations from a quartic minimum are contained in a
 $\mathcal{O}(\epsilon^{1/2}) \times \mathcal{O}(\epsilon^{2})$ box, and the
 whole curve is shifted by an $\mathcal{O}(\epsilon)$ amount.}
 \end{figure}

\subsection{Derivation of a generalised Ginzburg--Landau equation}
\label{subsec:derivation_GLE}

For the model PDE~\eqref{eq:PDE_nonlinear}, we consider only situations
where we have a small perturbation of the quartic marginal stability curve. At
the quartic minimum, an appropriate scaling is that if the solution $u$ is
of~$\mathcal{O}(\epsilon)$, then the slow time~$T$ and long length~$X$ should
be $\mathcal{O}(\epsilon^2)$ and $\mathcal{O}(\epsilon^{1/2})$ respectively:
the long length~$X$ is even longer than the $\mathcal{O}(\epsilon)$ scaling in
the case of a quadratic minimum in~\eqref{eq:GLE_scaling}. We therefore
restrict $f_{1}$ and~$f_{2}$ such that the two minima in the marginal stability
curve lie in an $\mathcal{O}(\epsilon^{1/2}) \times \mathcal{O}(\epsilon^{2})$
box, as illustrated in Figure~\ref{fig:marginal_scalings}. Other scalings are
possible.

To simplify the form of the marginal stability curve, we take $\mu_{p}=0$ and
consider the two minima and intermediate maximum of the function $\mu=p(k^2)$
in~\eqref{eq:polynomial}. The condition for an extremum is $p'(K)=0$, with
$K=k^2$: writing $\tfrac{1}{4}p'(K)$ as $(k^2-k_1^2)(k^2-k_2^2)(k^2-k_3^2)$,
with $k_1<k_2<k_3$, leads to the conclusion that
 \begin{equation}
 \label{eq:root_conditions}
 k_{1}^{2} + k_{2}^{2} + k_{3}^{2} = 3,
 \qquad
 k_{1}^{2} k_{2}^{2} + k_{1}^{2} k_{3}^{2} + k_{2}^{2} k_{3}^{2} = 3 + \tfrac{1}{2}f_{2},
 \qquad 
 k_{1}^{2} k_{2}^{2} k_{3}^{2} = 1 - \tfrac{1}{4}f_{1}.
 \end{equation}
To satisfy these three equations, 
we introduce new parameters $\gamma$ and~$\delta$, defined by
 \begin{equation}
 \label{eq:k_gamma_delta}
 k_{1}^{2} = 1 - \gamma - \tfrac{1}{2}\delta,
 \qquad
 k_{2}^{2} = 1 + 2 \gamma,
 \qquad 
 k_{3}^{2} = 1 - \gamma + \tfrac{1}{2}\delta,
 \end{equation}
such that the first equation in~\eqref{eq:root_conditions} is satisfied.
Requiring $k_1<k_2<k_3$ implies \hbox{$\delta>6|\gamma|\geq0$}.
Substituting~\eqref{eq:k_gamma_delta} into the second and third equations
in~\eqref{eq:root_conditions} yields
 \begin{align*}
 k_{1}^{2} k_{2}^{2} + k_{1}^{2} k_{3}^{2} + k_{2}^{2} k_{3}^{2} 
          &= 3 - 3\gamma^2 - \tfrac{1}{4}\delta^2
          = 3 + \tfrac{1}{2}f_{2}, \\
 k_{1}^{2} k_{2}^{2} k_{3}^{2} 
          &= 1 - 3\gamma^2 - \tfrac{1}{4}\delta^2 - \tfrac{1}{2}\gamma\delta^2 + 
             2\gamma^3
          = 1 - \tfrac{1}{4}f_{1}.
 \end{align*}
These are rearranged to result in expressions for $f_{1}$, $f_{2}$ and 
$f_{1}+2f_{2}$ in terms of $\gamma$ and~$\delta$:
 \begin{equation*}
 f_{1} = 12\gamma^2 + \delta^2 + 2\gamma\delta^2 - 8\gamma^3,
 \qquad
 f_{2} = -6\gamma^2 - \tfrac{1}{2}\delta^2,
 \qquad
 f_{1}+2f_{2} = 2\gamma\left(\delta^2 - 4\gamma^2\right).
 \end{equation*}
The restriction $\delta>6|\gamma|$ means that $\delta^2 - 4\gamma^2>0$,
and so $f_{1}+2f_{2}=0$ implies $\gamma=0$ and $k_2=1$.

Now, if the minima and maxima of the marginal stability curve are to satisfy the 
scalings in Figure~\ref{fig:marginal_scalings}, we need 
$\delta=\mathcal{O}(\epsilon^{1/2})$ and $\gamma$~no larger than this. We also 
need $p(k_{2}^2)-p(k_{1}^2)$ and $p(k_{2}^2)-p(k_{3}^2)$ to be 
$\mathcal{O}(\epsilon^2)$:
 \begin{equation*}
 p(k_{2}^2)-p(k_{1}^2) = \tfrac{1}{16}
                         \left(\delta-2\gamma\right)
                         \left(\delta+6\gamma\right)^3,
 \qquad
 p(k_{2}^2)-p(k_{3}^2) = \tfrac{1}{16}
                         \left(\delta+2\gamma\right)
                         \left(\delta-6\gamma\right)^3,
 \end{equation*}
so with $\delta=\mathcal{O}(\epsilon^{1/2})$ and $\gamma$~no larger, these
differences fit within~$\mathcal{O}(\epsilon^2)$. The overall marginal stability 
curve is shifted up by an amount 
$p(1)\approx\tfrac{1}{2}\left(12\gamma^2+\delta^2\right)=\mathcal{O}(\epsilon)$,
again within the scalings indicated in Figure~\ref{fig:marginal_scalings}. The
scaling of $\gamma$ and $\delta$ imply $f_{1}$ and $f_{2}$ are both
$\mathcal{O}(\epsilon)$, but that $f_{1}+2f_{2}=\mathcal{O}(\epsilon^{3/2})$. 

We are now in a position to perform the multiple scales analysis. The full set
of scalings used are
 \begin{align}
 \label{eq:wnlt_scaling}
 u &= \sum_{n=2}^{6} \epsilon^{n/2} u_{n/2}, 
 &
 \mu &= \epsilon \mu_{1} + \epsilon^{2} \mu_{2}, \nonumber \\
 \partial_{t} &= \epsilon^{2} \partial_{T},
 &
 f_{1} &= \epsilon f_{1},\\
 \partial_{x} &= \partial_{x} + \epsilon^{1/2} \partial_{X},
 &
 f_{2} &= \epsilon f_{2},
 \qquad\text{with $f_{1}+2f_{2}=\mathcal{O}(\epsilon^{3/2})$},
 \nonumber
 \end{align}
where $\epsilon\mu_{1}$ is the amount by which the marginal stability curve is
shifted. We also define a singular linear operator $\mathcal{L}$ to be 
$(1+\partial_{x}^2)^4$, with $\mathcal{L}e^{\pm ix}=0$.

Inserting these scalings into~\eqref{eq:PDE_nonlinear}, we obtain at leading
order $\mathcal{O}(\epsilon)$:
 \[
 0 = - \mathcal{L} u_{1},
 \]
which is satisfied by taking
 \[
 u_{1} = A(X,T) e^{i x} + \text{c.c.},
 \]
where \hbox{c.c.} represents the complex conjugate. It is useful to observe that
$\partial_{x}^2u_1=-u_1$.

Proceeding to $\mathcal{O}(\epsilon^{3/2})$, we have
 \[
 0 = - \mathcal{L} u_{3/2} - 8 \partial_{xX} \underbrace{(1 + \partial_{x}^{2})^{3} u_{1}}_{=0}.
 \]
The most convenient way to solve this is to set $u_{3/2} = 0$.

At $\mathcal{O}(\epsilon^{2})$, we have
 \[
 0 = \mu_{1} u_{1} - \mathcal{L} u_{2} 
    - 4 \partial_{X}^{2} \underbrace{\left( 1 + 9 \partial_{x}^{2} + 15 \partial_{x}^{4} + 7 \partial_{x}^{6} \right) u_{1}}_{=0} 
    + f_{1} \partial_{x}^{2} u_{1} 
    - f_{2} \partial_{x}^{4} u_{1} 
    + n_{2} u_{1}^{2}.
 \]
The terms proportional to $e^{ix}$ have a prefactor of $\mu_{1}-\left(f_{1}+f_{2}\right)$
We need to eliminate these terms in order to solve for $u_{2}$, which we do by 
setting $\mu_{1}  = f_{1} + f_{2}$. With this, the remaining terms are 
$0=-\mathcal{L} u_{2}+n_{2} u_{1}^{2}$, which can be solved to give
  \[
  u_{2} = n_{2} \left( \tfrac{1}{81}A^{2} e^{2ix} + \tfrac{1}{81}\bar{A}^{2} e^{-2ix} + 2 |A|^{2} \right).
 \]
The factors $\tfrac{1}{81}$ come from $\mathcal{L}e^{2ix}=81e^{2ix}$ when 
$\mathcal{L}$ is inverted.

At $\mathcal{O}(\epsilon^{5/2})$ we have
 \[
 0 = - \mathcal{L} u_{5/2} 
     - 8 \partial_{xX} (1 + \partial_{x}^{2})^{3} u_{2} 
     - 8 \partial_{x}^{3} \partial_{X} \underbrace{(3 + 10 \partial_{x}^{2} + 7 \partial_{x}^{4}) u_{1}}_{= 0} 
     + \underbrace{2 f_{1} \partial_{x} \partial_{X} u_{1} - 4 f_{2} \partial_{x}^{3} \partial_{X} u_{1}}_{=2(f_{1}+2f_{2})\partial_{xX}u_1}.
 \]
The only terms involving $e^{\pm i x}$ on the \hbox{RHS} are 
the two on the end that combine to have a prefactor of $f_{1}+2f_{2}$. As 
discussed above, this combination is a factor of $\epsilon^{1/2}$ smaller than 
$f_{1}$ and $f_{2}$ separately, so this term can be pushed 
to~$\mathcal{O}(\epsilon^{3})$ and dropped from this equation. With this, the 
two remaining terms in the equation can be solved for~$u_{5/2}$:
 \[
 u_{5/2} = \frac{32}{243}n_2\left(
                         i A A_{X} e^{2ix} - i \bar{A} \bar{A}_{X} e^{-2ix}
                         \right).
 \]
There is no constant term in~$u_{5/2}$.

Finally, continuing to $\mathcal{O}(\epsilon^{3})$, and including the term 
pushed from $\mathcal{O}(\epsilon^{5/2})$, we have
\begin{align*} 
 \partial_{T} u_{1} = \mu_{1} u_{2} 
 &  + \mu_{2} u_{1} - \mathcal{L} u_{3} 
    - 8 \partial_{xX} \left(1 + \partial_{x}^{2} \right)^{3} u_{5/2} \\
 &{}- 4 \partial_{X}^{2} \left( 1 + 9 \partial_{x}^{2} + 15 \partial_{x}^{4} + 7 \partial_{x}^{6} \right) u _{2} 
    - 2 \partial_{X}^{4} \left( 3 + 30 \partial_{x}^{2} + 35 \partial_{x}^{4} \right) u_{1} \\
 &{}+ f_{1} \partial_{x}^{2} u_{2} - f_{2} \partial_{x}^{4} u_{2} 
    + f_{1} \partial_{X}^{2} u_{1} - 6 f_{2} \partial_{x}^{2} \partial_{X}^{2} u_{1} 
    + 2 n_{2} u_{1} u_{2} + n_{3} u_{1}^{3}\\
 &{}+ 2(f_{1}+2f_{2})\partial_{xX}u_1.
 \end{align*}
The solvability condition requires the elimination of all terms proportional to 
$e^{\pm ix}$. There are no contributions from $\mathcal{L} u_{3}$ and from the 
terms linear in~$u_2$ and~$u_{5/2}$. Recalling that $\partial_{x}^{2}u_1=-u_1$,
the terms proportional to~$e^{ix}$ result in the solvability condition:
 \begin{equation}
 \label{eq:generalised_GLE}
 \frac{\partial A}{\partial T} = \mu_{2} A 
        + 2 i \nu_{1} \frac{\partial A}{\partial X} 
        + \nu_{2} \frac{\partial^{2} A}{\partial X^{2}} 
        - 16 \frac{\partial^{4} A}{\partial X^{4}} 
        + \left( 3 n_{3} + \tfrac{326}{81} n_{2}^{2} \right) |A|^{2} A,
 \end{equation}
where $\nu_{1} = f_{1} + 2 f_{2}$ and $\nu_{2} = f_{1} + 6 f_{2}$.
The nonlinear term is $n_A|A|^2A$, with $n_A$ defined in~\eqref{eq:nA}.
This is a generalisation of the Ginzburg--Landau equation~\eqref{eq:GLE},
and each term in the equation is, in terms of the original unscaled variables,
of order~$\mathcal{O}(\epsilon^{3})$.
Similar amplitude equations have been proposed before
\citep{Riley1989,Riecke1990,Raitt1995}, although not formally derived via
an asymptotic expansion, to model other problems with very flat marginal
stability curves.

\subsection{Nonlinear stability of rolls} 
\label{subsec:nonlinear_stability}

We can find roll solutions to~\eqref{eq:generalised_GLE} easily, and
use the equation to examine their stability to long wavelength Eckhaus
instabilities. We restrict ourselves to the supercritical case, where the
coefficient of the nonlinear term is negative. We consider a roll solution at
slightly off-critical wavenumber, \textit{i.e.},
 \begin{equation}
 \label{eq:GLE_offcritical_rolls}
  A(X,T) = R e^{i q X},
 \end{equation}
which corresponds to a solution $u = R \exp\left(i \left( 1 + \epsilon^{1/2} q
\right) x\right)$ of~\eqref{eq:PDE_nonlinear} since $X = \epsilon^{1/2} x$, and $q =
\mathcal{O}(1)$. These are also known as phase-winding solutions. 
Substituting~\eqref{eq:GLE_offcritical_rolls} into~\eqref{eq:generalised_GLE},
and rearranging, we obtain
 \begin{equation}
 \label{eq:wnlt_roll_amplitude}
 R^{2} 
       = - \frac{1}{n_A}\left( \mu_{2} - 2 \nu_{1} q - \nu_{2} q^{2} - 16 q^{4} \right),
 \end{equation}
where $n_A<0$ for a supercritical bifurcation.
The existence boundary of rolls, which is also the marginal stability curve,
is where $R^{2} = 0$, or equivalently
 \begin{equation}
 \label{eq:wnlt_marginal}
 \mu_{2} = 2 \nu_{1} q + \nu_{2} q^{2} + 16 q^{4},
 \end{equation}
with $\nu_1$ and $\nu_2$ here playing the roles of unfolding parameters
for the quartic minimum of the marginal stability curve.

To determine the stability of these roll solutions, we 
perturb~\eqref{eq:GLE_offcritical_rolls}, writing
 \begin{equation*}
 A(X,T) = R \left( 1 + r(X,T) \right) e^{i \left( q X + \phi(X,T) \right)},
 \end{equation*}
where $|r|$,$|\phi| \ll 1$, following \citet{Hoyle2006}. 
We substitute this expressions into~\eqref{eq:generalised_GLE},
linearise and separate the real and imaginary parts to obtain two linear 
constant coefficient PDEs for $r$ and~$\phi$. To solve these, we seek solutions 
of the form $e^{\sigma T + i m X}$, with $m\ll q$ and derive a quadratic
equation for the growth rate~$\sigma$. One root of this is always negative for a 
supercritical solution, and the other root is
 \begin{equation*}
  \sigma = - m^{2} \left( \nu_{2} + 96q^{2} + \frac{2}{n_A R^{2}} \left( \nu_{1} + \nu_{2} q + 32 q^{3} \right)^{2} \right) + \mathcal{O}(m^{3}).
 \end{equation*}
See \citet{Bentley2012} for details. 
 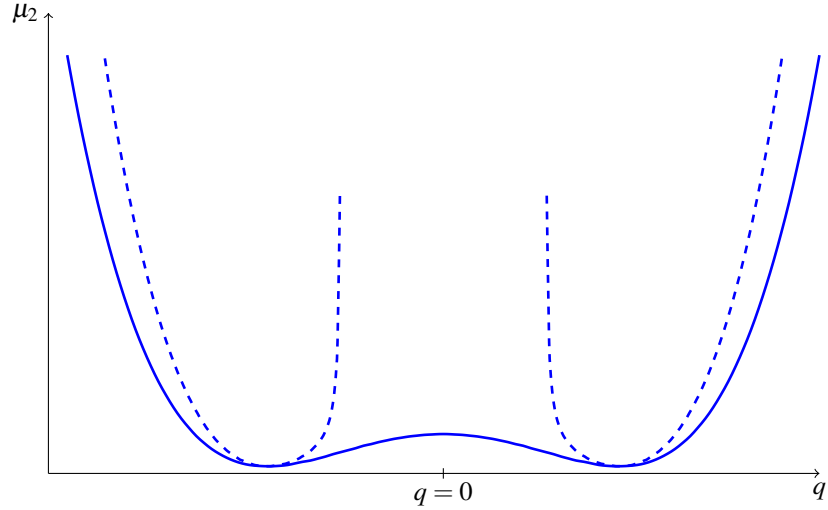
\begin{figure}
 \begin{center}
 \mbox{\beginpgfgraphicnamed{paper_figure_04}%
 \begin{tikzpicture}[xscale=25, yscale=350]
  \draw[line width=1pt,color=blue] plot[id=fig6_eck, domain=-0.2:0.2, samples=100, smooth]
          function{-0.28*x**2 + 16*x**4};
  \draw[line width=1pt,color=blue, dashed] plot[id=fig6_eck1, domain=-0.18:-0.055, samples=100, smooth]
          function{-0.28*x**2 + 16*x**4 + (-0.28*x + 32*x**3)**2 / (-0.28 + 96*x**2)};
  \draw[line width=1pt,color=blue, dashed] plot[id=fig6_eck2, domain=0.055:0.18, samples=100, smooth]
          function{-0.28*x**2 + 16*x**4 + (-0.28*x + 32*x**3)**2 / (-0.28 + 96*x**2)};
\draw[->] (-0.21,-0.0015) -- (0.2,-0.0015);
\draw[->] (-0.21,-0.0015) -- (-0.21,0.016);
\draw (0.2,-0.0015) node[below] {$q$};
\draw[-] (0,-0.0017) -- (0,-0.0013);
\draw (0,-0.0015) node[below] {$q=0$};
\draw (-0.21,0.016) node[left] {$\mu_{2}$};
\end{tikzpicture}
 \endpgfgraphicnamed}%
 \end{center}
 \caption{\label{fig:marg_and_eck} Marginal stability curve (solid) and Eckhaus
 curve (dashed) for $\nu_{1} = 0$ and $ \nu_{2} = -0.28$. The regions above the
 dashed curve are Eckhaus stable.}
\end{figure}
For the rolls to be unstable, we require $\sigma > 0$ for some $m$, 
giving a stability boundary defined by
 \[
 R^{2} = - 2 \frac{\left( \nu_{1} + \nu_{2} q + 32 q^{3} \right)^{2}}{n_A \left( \nu_{2} + 96 q^{2} \right)},
 \]
and, recalling~\eqref{eq:wnlt_roll_amplitude}, we find the Eckhaus stability
boundary
 \begin{equation}
 \label{eq:wnlt_eckhaus}
 \mu_{2} = 2 \nu_{1} q + \nu_{2} q^{2} + 16 q^{4} + \frac{2}{\nu_{2} + 96 q^{2}} \left( \nu_{1} + \nu_{2} q + 32 q^{3} \right)^{2}.
 \end{equation}
Figure~\ref{fig:marg_and_eck} shows an example of the
marginal~\eqref{eq:wnlt_marginal} and Eckhaus~\eqref{eq:wnlt_eckhaus} curves. Inside the
Eckhaus (dashed) curve, patterns are stable to long wavelength disturbances.
Note that $\nu_{2} < 0$ when there are two distinct minima, so~\eqref{eq:wnlt_eckhaus}
has two vertical asymptotes at $q = \pm \sqrt{-\nu_{2}/96}$.
Between these asymptotes, $\mu_2$~is below the marginal stability curve.

\subsection{The Proctor term}
\label{subsec:Proctor_term}

The presence of these asymptotes means that the Eckhaus boundary does not close
up in the middle, as might be expected. However, \citet{Proctor1991} considered
an amplitude equation similar to~\eqref{eq:generalised_GLE} but with the first
term on the RHS replaced by $\mu_2(1\pm{i}\partial_X)A$, and found that
(depending on parameters) the inner edges of the left and right stability
boundaries can meet, or the stability region can close with increasing~$\mu$
with two separate stability regions, and the stability boundaries can be
non-monotonic -- see \citet{Proctor1991} and \citet{Bentley2012} for examples.

However, when we considered the model
equation~\eqref{eq:PDE_nonlinear_Proctor}, with the Proctor term included, we
encountered difficulties. Writing $\mu=\epsilon\mu_1+\epsilon^2\mu_2$, as
in~\eqref{eq:wnlt_scaling}, turns out to be inadequate, so we modified the
scaling to include an additional
$\epsilon^{3/2}\mu_{3/2}\left(1+\partial_{x}^{2}\right)$ term, aiming to relate
$\mu_{3/2}$ to~$\mu_{p}$. It turns out that this doesn't work either, and
possibly further (potentially higher order) terms would need to be considered
for a consistent scaling \citep{Bentley2012}.

\subsection{The Lugiato--Lefever and complex Swift--Hohenberg equations}
\label{subsec:Kozyreff}

As an aside, we consider the 
Lugiato--Lefever equation
 \begin{equation}
 \label{eq:lugato_lefever}
 \frac{\partial \psi}{\partial t} = S - \left( 1 + i \bigtriangleup \right) \psi + i | \psi |^{2} \psi - i B_{2} \frac{\partial^{2} \psi}{\partial \tau^{2}} + i B_{4} \frac{\partial^{4} \psi}{\partial \tau^{4}},
 \end{equation}
which governs the envelope of the complex electromagnetic field~$\psi(t,\tau)$
inside a photonic crystal fibre cavity. In this equation, $S$~represents an
injected field, $\bigtriangleup$~is a cavity detuning, and $B_{2}$ and $B_{4}$
incorporate chromatic dispersion. The two time variables $t$ and $\tau$
represent, respectively, the average evolution of~$\psi$ over one cavity round
trip and the fast variations of~$\psi$. We note the similarities
between~\eqref{eq:lugato_lefever} and~\eqref{eq:PDE_nonlinear}: the independent
variable~$\tau$ is equivalent to~$x$, there is a cubic nonlinearity, the
inhomogeneous $S$ term breaks the $\psi \rightarrow -\psi$ symmetry and implies
a quadratic nonlinearity, and a fourth order complex equation is equivalent to
an eighth order real equation.

This system also allows marginal stability curves with a double minimum,
and \citet{Kozyreff2009} derived a similar amplitude equation for this case.
Their
analysis concentrates only on the case when the two minima
occur at the same height, equivalent to $\gamma=0$ in~\eqref{eq:k_gamma_delta}. 
This is a degenerate situation however; to recover the
generic situation it is necessary to include a $B_{3} \partial^{3} \psi /
\partial \tau^{3}$ term in~\eqref{eq:lugato_lefever}. The degenerate case is
chosen by \citet{Kozyreff2009} both as a means of simplifying the analysis and as a
situation easily achievable experimentally. 

With $\gamma=0$, \citet{Bentley2012} showed that the appropriate amplitude
equation for the Lugiato--Lefever equation is~\eqref{eq:generalised_GLE} but with
$\nu_1=0$. This special case is interesting: with $\nu_1=0$ and for $\nu_2<0$,
equation~\eqref{eq:generalised_GLE} is equivalent to the \emph{complex} Swift--Hohenberg
equation:
 \begin{equation}
 \label{eq:complex_SHE}
 \frac{\partial A}{\partial T} = \lambda A - \left( 1 + \frac{\partial^{2}}{\partial X^{2}} \right)^{2} A + n_{A} |A|^{2} A, 
 \end{equation}
having scaled and changed notation. This differs from the \emph{real} 
SH~equation~\eqref{eq:SHE} in that there is no quadratic term 
and that the cubic nonlinearity is $|A|^2A$ rather than~$u^3$.
It also differs from the more usual complex SH equation, 
which has complex coefficients \citep{Sakaguchi1997}. 
The complex SH~equation with real coefficients has been investigated by 
\citet{Gelens2010} and is equivalent to to the equation studied by
\citet{Raitt1995}.
However, \citet{Kozyreff2009} find the real SH~equation~\eqref{eq:SHE} as the
amplitude equation for the Lugiato--Lefever problem, rather than the complex
Swift--Hohenberg equation with real coefficients~\eqref{eq:complex_SHE}. We
believe this is an error, and it means that the interpretation by
\citet{Kozyreff2009} of localised solutions of~\eqref{eq:lugato_lefever} as
localised solution of~\eqref{eq:SHE} is incorrect; instead they should be
interpreted in terms of localised solutions of~\eqref{eq:complex_SHE}, which are
different. We investigate these localised solutions in more detail below.


\section{Lyapunov functional and the first integral}
\label{sec:Lyapunov_first_int}

In the Swift--Hohenberg equation, the Lyapunov functional and first integral
are useful for finding localised solutions: if two solutions are to be
connected by a stationary front, they should have the same values of the first
integral and similar values of the Lyapunov functional. In this section, we
generalise the SH results to the model equation~\eqref{eq:PDE_nonlinear}.

Multiplying the steady version of~\eqref{eq:PDE_nonlinear} by $u_x$ and
integrating by parts gives us a first integral:
 \begin{align*}
 \mathcal{H} &= - \tfrac{1}{2} \left( \mu - 1 \right) u^{2} 
                - \tfrac{1}{3}n_{2}u^{3} 
                - \tfrac{1}{4}n_{3}u^{4} 
                + \tfrac{1}{2} \left( 4 - f_{1} \right) u_{x}^{2}
                - \tfrac{1}{2} \left( 6 + f_{2} \right) \left( u_{xx}^{2} - 2 u_{x} u_{xxx} \right) \\ \nonumber
    & \qquad {} + 2 u_{xxx}^{2} 
                - 4 u_{xx} u_{xxxx}
                + 4 u_{x} u_{xxxxx} 
                - \tfrac{1}{2} u_{xxxx}^{2} 
                + u_{xxx} u_{xxxxx} 
                - u_{xx} u_{xxxxxx} 
                + u_{x} u_{xxxxxxx}.
 \end{align*}
Any steady solution of~\eqref{eq:PDE_nonlinear} must have
$\frac{d\mathcal{H}}{dx}=0$. It is a straight-forward modification to include
the Proctor term from~\eqref{eq:PDE_nonlinear_Proctor}.

Using the Lyapunov functional for the Swift--Hohenberg
equation~\eqref{eq:SH_Lyaponov} as a starting point, we define a similar
functional for~\eqref{eq:PDE_nonlinear}, namely
 \begin{equation*}
 \mathcal{F}[u] = \int_{0}^{L} \left( 
                       - \tfrac{1}{2} \mu u^{2} 
                       + \tfrac{1}{2} \left( \left( 1 + \partial_{x}^{2} \right)^{2} u \right)^{2} 
                       + \tfrac{1}{2} f_{1} \left( \partial_{x} u \right)^{2}
                       + \tfrac{1}{2} f_{2} \left( \partial_{x}^{2} u \right)^{2} 
                       - \tfrac{1}{3} n_{2}u^{3} 
                       - \tfrac{1}{4} n_{3}u^{4} \right) dx.
 \end{equation*}
One can readily show that 
 \[
 \frac{\partial u}{\partial t} = - \frac{\delta \mathcal{F}}{\delta u}
 \qquad \text{and} \qquad
 \frac{d \mathcal{F}}{d t} = - \int_{0}^{L} \left( \frac{\partial u}{\partial t} \right)^{2} dx \leq 0.
 \]
It is possible to show that $\mathcal{F}[u]$ is bounded below provided $f_2>-4$ 
and~$n_3<0$
\citep{Bentley2012}, 
and so, as in the SH equation, that stable solutions are local minima
of~$\mathcal{F}[u]$. Similarly, it is a straight-forward modification to
include the Proctor term from~\eqref{eq:PDE_nonlinear_Proctor}.

\begin{figure}
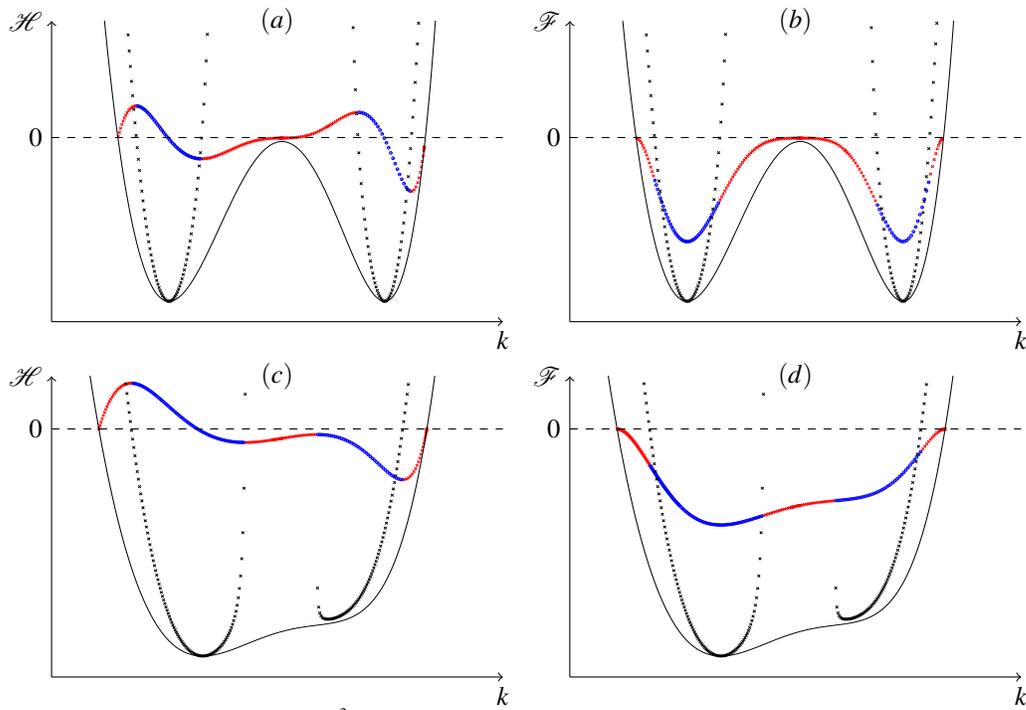

 \begin{center}
 \mbox{\beginpgfgraphicnamed{paper_figure_05ab}%
 \begin{tikzpicture}
 \node [inner sep=0pt,outer sep=0pt] at (0,0) {\includegraphics[trim=5mm 15mm 10mm 10mm, clip, width=6.0cm, height=4cm]{ch6_1_hm1.pdf}};
\draw[->] (-3.0,-2.0) -- (3.0,-2.0);
\draw[->] (-3.0,-2.0) -- (-3.0,2.0);
\draw (0,2.0) node(above) {$(a)$};
\draw (-3.0,2.0) node[left] {$\mathcal{H}$};
\draw ( 3.0,-2.0) node[below] {$k$};
\draw[dashed] (-3.0,0.45) -- (3.0,0.45);
\draw (-3.0,0.45) node[left] {$0$};
  \end{tikzpicture}
  \begin{tikzpicture}
  \node [inner sep=0pt,outer sep=0pt] at (0,0) {\includegraphics[trim=5mm 15mm 10mm 10mm, clip, width=6.0cm, height=4cm]{ch6_1_lf1.pdf}};
\draw[->] (-3.0,-2.0) -- (3.0,-2.0);
\draw[->] (-3.0,-2.0) -- (-3.0,2.0);
\draw (0,2.0) node(above) {$(b)$};
\draw (-3.0,2.0) node[left] {$\mathcal{F}$};
\draw ( 3.0,-2.0) node[below] {$k$};
\draw[dashed] (-3.0,0.45) -- (3.0,0.45);
\draw (-3.0,0.45) node[left] {$0$};
\end{tikzpicture}
\endpgfgraphicnamed}%
 \end{center}
  \vspace{-5ex}
 \begin{center}
 \mbox{\beginpgfgraphicnamed{paper_figure_05cd}%
 \begin{tikzpicture}
 \node [inner sep=0pt,outer sep=0pt] at (0,0) {\includegraphics[trim=5mm 15mm 10mm 5mm, clip, width=6.0cm, height=4cm]{ch6_3_hm1.pdf}};
\draw[->] (-3.0,-2.0) -- (3.0,-2.0);
\draw[->] (-3.0,-2.0) -- (-3.0,2.0);
\draw (0,2.0) node(above) {$(c)$};
\draw (-3.0,2.0) node[left] {$\mathcal{H}$};
\draw ( 3.0,-2.0) node[below] {$k$};
\draw[dashed] (-3.0,1.30) -- (3.0,1.30);
\draw (-3.0,1.30) node[left] {$0$};
  \end{tikzpicture}
  \begin{tikzpicture}
  \node [inner sep=0pt,outer sep=0pt] at (0,0) {\includegraphics[trim=5mm 15mm 10mm 5mm, clip, width=6.0cm, height=4cm]{ch6_3_lf1.pdf}};
\draw[->] (-3.0,-2.0) -- (3.0,-2.0);
\draw[->] (-3.0,-2.0) -- (-3.0,2.0);
\draw (0,2.0) node(above) {$(d)$};
\draw (-3.0,2.0) node[left] {$\mathcal{F}$};
\draw ( 3.0,-2.0) node[below] {$k$};
\draw[dashed] (-3.0,1.30) -- (3.0,1.30);
\draw (-3.0,1.30) node[left] {$0$};
\end{tikzpicture}
\endpgfgraphicnamed}%
 \end{center}
  \vspace{-5ex}
 \caption{\label{fig:H_and_F_vs_k} Plot of $(a)$ $30 \mathcal{H}$ and $(b)$ $3
 \times 10^{3} \mathcal{F}$, against $k$, for $f_{1}=0.14$, $f_{2}=-0.07$ and
 $\mu=0.07$, computed using AUTO. 
 $(c,d)$: $15 \mathcal{H}$ and $0.5\times10^{3}\mathcal{F}$, 
 for $f_{1}=0.0488$, $f_{2}=-0.0227$ and $\mu=0.0306$. 
 The red crosses correspond to Eckhaus unstable
 wavenumbers, and the blue circles to Eckhaus stable wavenumbers. The marginal
 stability curve (solid black) and Eckhaus curves (black crosses) are also shown
 -- the change from red crosses to blue circles does not exactly match the
 Eckhaus boundary owing to the scaling of $\mathcal{H}$ and~$\mathcal{F}$.
 Note that in $(c,d)$ there is only one minimum in the marginal stability
 curve, but there is a region of Eckhaus-stable patterns above the
 almost-minimum.}
 \end{figure}

Small-amplitude solutions of~\eqref{eq:PDE_nonlinear} can be found using the
generalised GL equation~\eqref{eq:generalised_GLE}, and these can be used to
find weakly nonlinear estimates of $\mathcal{H}$ and~$\mathcal{F}$.
Alternatively, fully nonlinear solutions of~\eqref{eq:PDE_nonlinear}  can be
found numerically, for example by using AUTO \citep{Doedel2007a}. Examples of
$\mathcal{H}$ and~$\mathcal{F}$ computed numerically in this way are shown in
Figure~\ref{fig:H_and_F_vs_k}, in the cases where the minima in the marginal
stability curve are at the same height, and where there is a single minimum
with an almost-minimum just outside the cusp. These were computed using an
initial solution on a domain of size $L=2\pi$, which was then continued in~$L$,
increasing and decreasing to cover the range of wavenumbers for which a pattern
solution exists. We note that $\mathcal{H}$ and~$\mathcal{F}$ are both zero at
the extremities of the existence region, and that the extrema of~$\mathcal{H}$
correspond to the Eckhaus stability boundaries.

The third tool, mentioned in $\S$\ref{sec:introduction}, is the normal form of
this variant of the Hamiltonian--Hopf bifurcation with four-fold degenerate
eigenvalues~$\pm{i}$. We derive the normal form for this bifurcation in
Appendix~\ref{sec:appendix} (see equations~\eqref{eq:app:normal_form_version_1}
and~\eqref{eq:app:normal_form_version_2}), but as we haven't found any first integrals of the
normal form, we don't see a way to use it at this point.

\section{Localised solutions}
\label{sec:localised}

The first integral~$\mathcal{H}$ and the Lyapunov functional~$\mathcal{F}$ are
two imporant tools for identifying where a pattern of one type can be localised
within a background of a pattern of another type: the values of the first
integral for the two patterns must be the same (since
$\frac{d\mathcal{H}}{dx}=0$ on any steady solution) and the Lyapunov functional
for the two patterns should be approximately the same.

We see from the example in Figure~\ref{fig:H_and_F_vs_k}$(a,b)$ that there is a
range of possible wavenumbers which satisfy the requisite criteria for patterns
with two different wavenumbers coexisting, namely there are wavenumbers that
have the same value of $\mathcal{H}$, and there are (different) wavenumbers with
the same value of $\mathcal{F}$. To narrow down the allowable wavenumbers, we
look for wavenumber pairs $(k_{-} < 1, k_{+} > 1)$ such that $\mathcal{H}(k_{-})
= \mathcal{H}(k_{+})$ and $\mathcal{F}(k_{-}) = \mathcal{F}(k_{+})$. We do this
by looking for intersections of contour lines plotted in the $(k_{+},k_{-})$
plane. For the parameter values in Figure~\ref{fig:H_and_F_vs_k}$(a,b)$, a pair of
wavenumbers that satisfy this condition are $(k_{+},k_{-})=(1.0778,0.8839)$. We
view such a point as an extension of the Maxwell point for the Swift--Hohenberg
equation, though it plays a different role: in the Swift--Hohenberg the
localised solutions are organised about the Maxwell point, whereas we use the
extension merely as a starting point to look for localised solutions.
 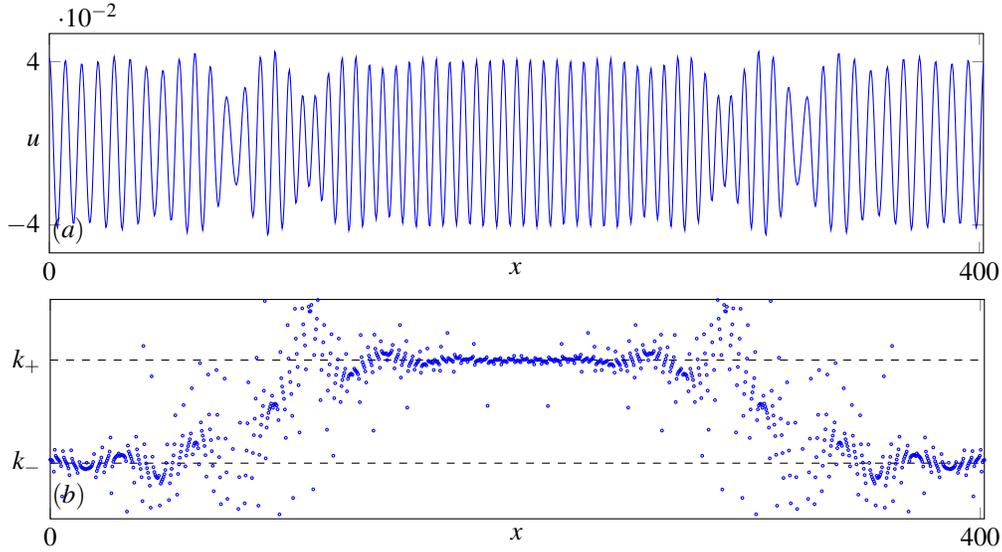
\begin{figure}
 \begin{center}
\begin{tabular}{c}
 \mbox{\beginpgfgraphicnamed{paper_figure_06a}%
  \begin{tikzpicture}
\begin{axis}[width=14cm,height=4.5cm,xtick={0,400},ytick={-0.04,0.04},xmin=0, xmax=401.7312]
 \addplot[mark=none, color=blue] table[x index=0, y index=1]{\tbl};
\end{axis}
\draw (6.2,0) node[below] {$x$};
\draw (0,1.5) node[left] {\hspace{1.5em} $u$};
\draw (0.25,0) node[above] {$(a)$};
\end{tikzpicture}
\endpgfgraphicnamed}%
\\
 \mbox{\beginpgfgraphicnamed{paper_figure_06b}%
  \begin{tikzpicture}
\begin{axis}[width=14cm,height=4.5cm,xtick={0,400},ytick={-0.04,0.04},scaled ticks=false,ymin=0.8, ymax=1.2, xmin=0, xmax=401.7312]
 \addplot[color=blue,mark=o,only marks,mark size=0.5pt] table[x index=0, y index=2]{\tbl};
\addplot[mark=none,dashed] coordinates{(0,0.9015) (401.7312,0.9015)};
\addplot[mark=none,dashed] coordinates{(0,1.0895) (401.7312,1.0895)};
\end{axis}
\draw (6.2,0) node[below] {$x$};
\draw (0,0.75) node[left] {\hspace{1em} $k_{-}$};
\draw (0,2.1) node[left] {$k_{+}$};
\draw (0.25,0) node[above] {$(b)$};
\end{tikzpicture}
\endpgfgraphicnamed}%
  \end{tabular}
 \end{center}
  \vspace{-2ex}
\caption{\label{fig:PDE_example} Numerical simulation of mixed pattern initial
condition with wavenumbers $k=58/64=0.90625$ and $k=70/64=1.09375$. The
parameter values are: $f_{1}=0.14$, $f_{2}=-0.07$ and $\mu = 0.07$,
$n_{2}=0.1$, $n_{3}=-1$, $L=64\times2\pi$ and timestep~$0.01$. $(a)$~final
solution profile $u(x)$. $(b)$~approximation to the local wavenumber, defined
in \eqref{eq:local_wavenumber}. The values indicated are $k_{-}=0.9015$ and
$k_{+}=1.0895$.}
 \end{figure}

On a periodic domain of length~$L$, wavenumbers are restricted to  integer
multiples of $k=2\pi/L$. We therefore construct an initial condition consisting
of a region of pattern with wavenumber close to~$1.0778$ embedded in a
background of pattern with wavenumber close to~$0.8839$. Fixing a domain size
$L=64\times2\pi$, we choose $k_{-} = 58/64$ and $k_{+} = 70/64$, and
solve~\eqref{eq:PDE_nonlinear} using a second-order numerical scheme based on
Exponential Time Differencing \citep{Cox2002}. One example solution after
transients can be seen in Figure~\ref{fig:PDE_example}, which demonstrates that
localised solutions to~\eqref{eq:PDE_nonlinear} exist and are stable. This
solution is made up of a high wavenumber patch ($k\approx1.0895$) in the
centre of the domain, surrounded by low wavenumber regions ($k\approx0.9015$).
The approximation to the local wavenumber in Figure~\ref{fig:PDE_example}$(b)$
is found via
 \begin{equation}
 \label{eq:local_wavenumber}
 \textrm{local wavenumber} = \sqrt{\frac{- u_{xx}}{u}},
 \end{equation}
and matches (at least approximately) the expected values. The oscillations seen
in the amplitude and in the local wavenumber represent a beating between the
two constituent wavenumbers $k_{-}$ and~$k_{+}$ as the pattern adjusts from one
wavenumber to the other.

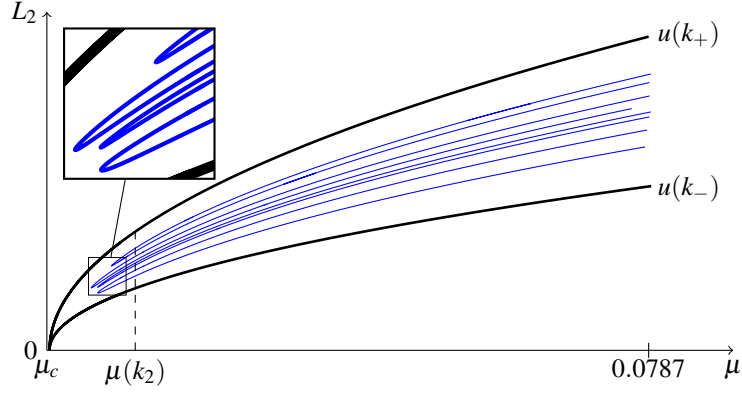
\begin{figure}
 \begin{center}
 \mbox{\beginpgfgraphicnamed{paper_figure_07}%
\begin{tikzpicture}[spy using outlines={rectangle, magnification=4,connect spies}]
\begin{axis}[axis lines=none,xtick=\empty,ytick=\empty,scaled ticks=false,width=10cm,height=6cm,xmin=-0.0005,xmax=0.01,ymin=0,ymax=0.35]
 \addplot[restrict x to domain=0:0.01,mark=none,color=blue] table[x index=0,y index=1]{b0.dat};
 \addplot[restrict x to domain=0:0.01,mark=none,color=black,line width=1.0pt] table[x index=0,y index=1]{b1.dat};
 \addplot[restrict x to domain=0:0.01,mark=none,color=black,line width=1.0pt] table[x index=0,y index=1]{b14.dat};
 \addplot[restrict x to domain=0:0.01,mark=none,color=blue] table[x index=0,y index=1]{b2.dat};
 \addplot[restrict x to domain=0:0.01,mark=none,color=blue] table[x index=0,y index=1]{b3.dat};
 \addplot[restrict x to domain=0:0.01,mark=none,color=blue] table[x index=0,y index=1]{b4.dat};
 \addplot[restrict x to domain=0:0.01,mark=none,color=blue] table[x index=0,y index=1]{b5.dat};
 \addplot[restrict x to domain=0:0.01,mark=none,color=blue] table[x index=0,y index=1]{b6.dat};
 \addplot[restrict x to domain=0:0.01,mark=none,color=blue] table[x index=0,y index=1]{b7.dat};
 \addplot[restrict x to domain=0:0.01,mark=none,color=blue] table[x index=0,y index=1]{b8.dat};
 \addplot[restrict x to domain=0:0.01,mark=none,color=blue] table[x index=0,y index=1]{b9.dat};
 \addplot[restrict x to domain=0:0.01,mark=none,color=blue] table[x index=0,y index=1]{b10.dat};
 \addplot[restrict x to domain=0:0.01,mark=none,color=blue] table[x index=0,y index=1]{b11.dat};
 \addplot[restrict x to domain=0:0.01,mark=none,color=blue] table[x index=0,y index=1]{b12.dat};
 \addplot[restrict x to domain=0:0.01,mark=none,color=blue] table[x index=0,y index=1]{b13.dat};
 \addplot[restrict x to domain=0:0.01,mark=none,color=blue] table[x index=0,y index=1]{b15.dat};
 \addplot[restrict x to domain=0:0.01,mark=none,color=blue] table[x index=0,y index=1]{b16.dat};
 \addplot[restrict x to domain=0:0.01,mark=none,color=blue] table[x index=0,y index=1]{b17.dat}; 
\coordinate (spypoint) at (axis cs:0.00097,0.078);
\coordinate (spyviewer) at (axis cs:0.0015,0.26);
\spy[width=2cm,height=2cm] on (spypoint) in node [fill=white] at (spyviewer);
\end{axis}
 \draw[->] (0.375,0) -- (0.375,4.5);
 \draw[->] (0.375,0) -- (9.5,0);
 \draw[dashed] (1.55,0) -- (1.55,1.6);
 \draw (9.5,0) node[below] {$\mu$};
 \draw (0.375,0) node[below] {$\mu_{c}$};
 \draw (0.375,0) node[left] {$0$};
 \draw (8.375,0) node[below] {$0.0787$};
 \draw (8.375,-0.1) -- (8.375,0.1);
 \draw (0.375,4.5) node[left] {$L_{2}$};
 \draw (8.375,4.2) node[right] {$u(k_{+})$};
 \draw (8.375,2.15) node[right] {$u(k_{-})$};
 \draw (1.55,0) node[below] {$\mu(k_{2})$};
 \end{tikzpicture}
 \endpgfgraphicnamed}%
 \end{center}
 \caption{\label{fig:bifurcation_diag} Bifurcation diagram for parameter values
 $f_{1}=0.14$, $f_{2}=-0.07$. The critical value of the bifurcation parameter
 $\mu$ is $\mu_{c}=0.0687$, and the local maximum of the marginal stability
 curve occurs at $(k_{2}, \mu(k_{2})) = (1,0.0699)$. The thick black branches
 correspond to periodic patterns with wavenumbers $k_{-}$, $k_{+}$, and are
 given for reference. The blue branches correspond to localised solutions. Inset
 is a magnification of the saddle-nodes.}
 \end{figure}

We can continue this and other solutions we have found in AUTO, continuing
in~$\mu$ to obtain the bifurcation diagram shown in Figure~\ref{fig:bifurcation_diag}.
Localised solutions lie on distinct branches that do not join up, created in
saddle-node bifurcations. The solution branches corresponding to a pattern of
single wavelength $k_{-}$ or $k_{+}$ are included for reference. On localised
solutions branches closer to the $k_{-}$ branch, more of the domain is filled by
the $k_{-}$ pattern than the $k_{+}$ pattern, and \textit{vice versa}.
The localised solution branches extend to values of $\mu$ below the value at the
local maximum of the marginal stability curve~($\mu(k_2)=0.0699$).

We find similar disconnected branches of localised solutions even when the local
minima in the marginal stability curve are not at the same height, and even just
outside the cusp, where there is a single minimum and a second almost-minimum,
as in Figure~\ref{fig:H_and_F_vs_k}$(c,d)$. Solutions in this region rely on there
being a region of Eckhaus-stable patterns still present above the
almost-minimum, disconnected from the marginal stability curve.

\subsection{Interpretation of localised solutions via the amplitude equation}

The amplitude equation~\eqref{eq:generalised_GLE} has phase-winding solutions
$A=Re^{iqX}$, with $R$ and~$q$ related by~\eqref{eq:wnlt_roll_amplitude}.
Previous work on this amplitude equation with $\nu_1=0$ \citep{Raitt1995,
Raitt1997, Gelens2009, Gelens2010} -- the complex 
Swift--Hohenberg equation with real coefficients -- has identified solutions 
that are combinations of two phase-winding
solutions with positive and negative values of~$q$: these are precisely the 
localised patterns we found in the model PDE~\eqref{eq:PDE_nonlinear} and shown in 
Figure~\ref{fig:PDE_example} (with $\nu_1=f_{1}+2f_{2}=0$). Here we extend this
interpretation to the case~\hbox{$\nu_1\neq0$}.

In order to find localised solutions, we could develop a first integral and a 
Lyapunov function for~\eqref{eq:generalised_GLE} and look for pairs of solutions with the 
same values of the quantities. We reserve this for future work, and instead 
locate localised solutions of~\eqref{eq:generalised_GLE} by starting with a mixture of 
two phase-winding solutions with constituent wavenumbers $q_{-}$ and~$q_{+}$, 
and timestepping the~\hbox{PDE}. Two examples are shown in 
Figures~\ref{fig:EGL_example_1} (with $\nu_1\neq0$ and a small value of~$\nu_2$)
and~\ref{fig:EGL_example_2} (with $\nu_1=0$ and a larger value of~$\nu_2$).
 \begin{figure}
 \begin{center}
  \mbox{\beginpgfgraphicnamed{paper_figure_08}%
  \begin{tabular}{cc}
  \begin{tikzpicture}
   \begin{axis}[width=7cm, height=4.5cm,xtick={0, 400},ytick={-0.04,0.04},xmin=0,xmax=401.12]
    \addplot[color=blue,mark=none] table[x index=0,y index=1]{ch6_amp_004.dat};
    \addplot[color=blue,dashed,mark=none] table[x index=0,y index=2]{ch6_amp_004.dat}; 
   \end{axis}
   \draw (0.3,0) node[above] {$(a)$};
   \draw (2.8,0) node[below] {$x$};
   \draw (0,1.55) node[left] {$A$};
  \end{tikzpicture}
 &
  \begin{tikzpicture}
   \begin{axis}[width=7cm, height=4.5cm,xtick={0, 400},ytick={0.035,0.038},xmin=0,xmax=401.12]
    \addplot[color=blue,mark=none] table[x index=0,y index=3]{ch6_amp_004.dat};
   \end{axis}
   \draw (0.3,0) node[above] {$(b)$};
   \draw (2.8,0) node[below] {$x$};
   \draw (0,1.55) node[left] {$|A|$};
  \end{tikzpicture}
\\
  \begin{tikzpicture}
   \begin{axis}[width=7cm, height=4.5cm,xtick={0, 400},ytick={-0.06,0.06},xmin=0,xmax=401.12]
    \addplot[color=blue,mark=none] table[x index=0,y index=4]{ch6_amp_004.dat};
   \end{axis}
   \draw (0.3,0) node[above] {$(c)$};
   \draw (2.8,0) node[below] {$x$};
   \draw (0,1.55) node[left] {$u$};
  \end{tikzpicture}
&
  \begin{tikzpicture}
   \begin{axis}[width=7cm, height=4.5cm,xtick={0, 400},ytick={0.9,1.1},xmin=0,xmax=401.12,ymin=0.9,ymax=1.1]
    \addplot[color=blue,mark=o,only marks,mark size=0.5pt] table[x index=0,y index=5]{ch6_amp_004.dat};
    \draw[dashed] (axis cs:0,0.946) -- (axis cs:401.12,0.946);
    \draw[dashed] (axis cs:0,1.051) -- (axis cs:401.12,1.051);
   \end{axis}
   \draw (0.3,0) node[above] {$(d)$};
   \draw (2.8,0) node[below] {$x$};
   \draw (0,1.55) node[left] {$k$};
   \draw (0,2.15) node[left] {$k_{+}$};
   \draw (0,0.65) node[left] {$k_{-}$};
  \end{tikzpicture}
  \end{tabular}
  \endpgfgraphicnamed}%
 \end{center}
 \caption{\label{fig:EGL_example_1} Solution to~\eqref{eq:generalised_GLE}, with $\nu_{1}=0.0034$, $\nu_{2}=-0.0874$, $\mu_{2}=0.004$, $n_{2}=0.1$ and $n_{3}=-1$. $(a)$: the real (solid) and imaginary (dashed) parts of the amplitude $A$, $(b)$: the absolute value $|A|$, $(c)$: the reconstructed solution $u = A e^{i x} + \bar{A} e^{-i x}$, and $(d)$: an approximation to the local wavenumber of the reconstructed solution in $(c)$. The dashed lines correspond to $(k_{-},k_{+}) = (0.946,1.053)$.}
 \end{figure}
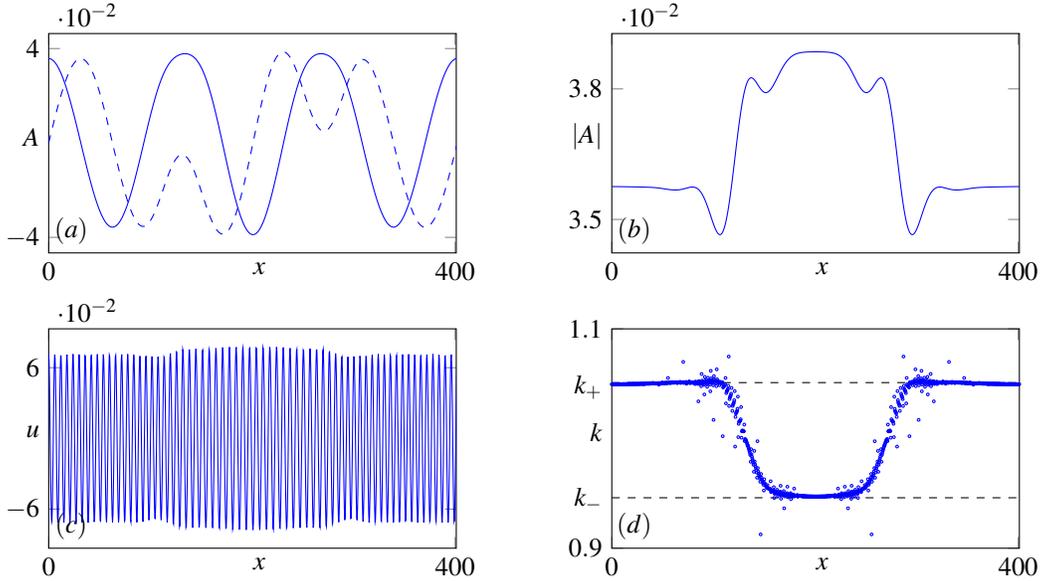
 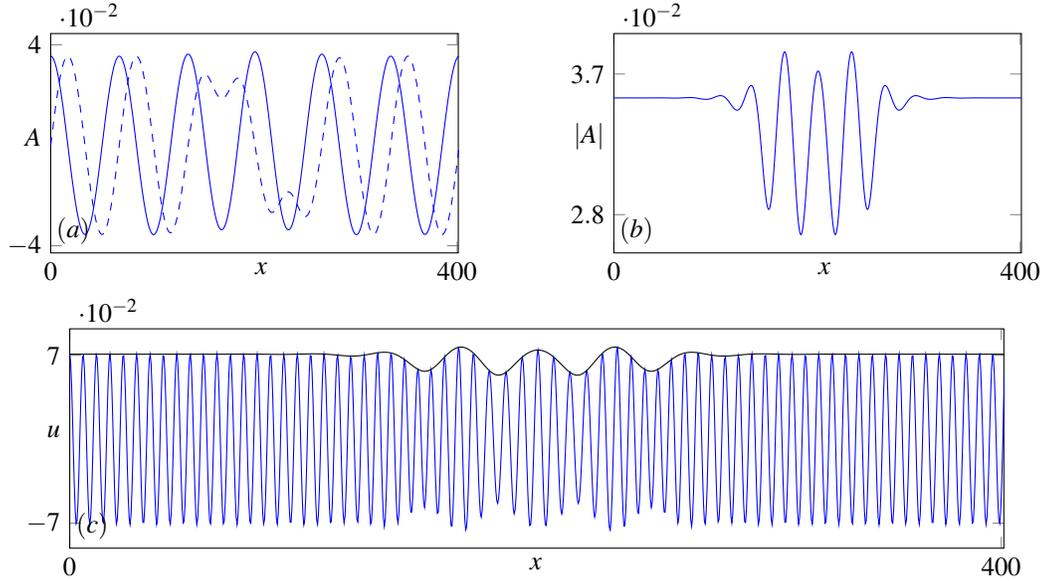
\begin{figure}
 \begin{center}
 \mbox{\beginpgfgraphicnamed{paper_figure_09}%
 \begin{tabular}{c}
  \begin{tabular}{cc}
  \begin{tikzpicture}
   \begin{axis}[width=7cm, height=4.5cm,xtick={0, 400},ytick={-0.04,0.04},xmin=0,xmax=401.12]
    \addplot[color=blue,mark=none] table[x index=0,y index=1]{ch6_amp_0025.dat};
    \addplot[color=blue,dashed,mark=none] table[x index=0,y index=2]{ch6_amp_0025.dat}; 
   \end{axis}
   \draw (0.3,0) node[above] {$(a)$};
   \draw (2.8,0) node[below] {$x$};
   \draw (0,1.55) node[left] {$A$};
  \end{tikzpicture}
&
  \begin{tikzpicture}
   \begin{axis}[width=7cm, height=4.5cm,xtick={0, 400},ytick={0.028,0.037},xmin=0,xmax=401.12]
    \addplot[color=blue,mark=none] table[x index=0,y index=3]{ch6_amp_0025.dat};
   \end{axis}
   \draw (0.3,0) node[above] {$(b)$};
   \draw (2.8,0) node[below] {$x$};
   \draw (0,1.55) node[left] {$|A|$};
  \end{tikzpicture}
  \end{tabular}
\\
  \begin{tikzpicture}
   \begin{axis}[width=14cm, height=4.5cm,xtick={0, 400},ytick={-0.07,0.07},xmin=0,xmax=401.12]
    \addplot[color=blue,mark=none] table[x index=0,y index=4]{ch6_amp_0025.dat};
    \addplot[color=black,mark=none] table[x index=0,y expr=2*\thisrowno{3}]{ch6_amp_0025.dat};
   \end{axis}
   \draw (0.3,0) node[above] {$(c)$};
   \draw (6.2,0) node[below] {$x$};
   \draw (0,1.55) node[left] {$u$};
  \end{tikzpicture}
  \end{tabular}
 \endpgfgraphicnamed}%
 \end{center}
 \caption{\label{fig:EGL_example_2} Solution to~\eqref{eq:generalised_GLE}, with $\nu_{1}=0$, $\nu_{2}=-0.28$, $\mu_{2}=0.0025$, $n_{2}=0.1$ and $n_{3}=-1$. $(a)$: the real (solid) and imaginary (dashed) parts of the amplitude $A$, $(b)$: the absolute value $|A|$, and $(c)$: the reconstructed solution $u = A e^{i x} + \bar{A} e^{-i x}$, with the amplitude $2 |A|$ plotted also.}
\end{figure}
Figure~\ref{fig:EGL_example_1}$(a)$ shows the real (solid) and imaginary (dashed)
parts of the solution $A$, $\Re(A)$ and $\Im(A)$ respectively. We can clearly
see the transition from $q_{-}$ to $q_{+}$ from $\Im(A)$, as the peaks of
$\Im(A)$ shift from being on the right of the peaks of $\Re(A)$ to the left, and
then back again. Figure~\ref{fig:EGL_example_1}$(b)$ shows~$|A|$, (half) the
amplitude of the reconstructed pattern $ u = A e^{i x} + \bar{A} e^{-i x}$ shown
in Figure~\ref{fig:EGL_example_1}$(c)$. Figure~\ref{fig:EGL_example_1}$(d)$ shows the
approximation to the local wavenumber of the reconstructed pattern. 

The example in Figure~\ref{fig:EGL_example_2}, with~$\nu_1=0$, is similar, but 
shows more pronounced beating between the two wavenumbers, evident in the 
reconstructed solution and in Figure~\ref{fig:PDE_example}. In this case, 
the localised patterns found in the model PDE are seen to be interpreted 
correctly in terms of localised combinations of phase-winding solutions of the 
complex Swift--Hohenberg equation, rather than as localised solutions of the 
real Swift--Hohenberg equation, as suggested by \citet{Kozyreff2009}. The complex 
Swift--Hohenberg equation does (since the coefficients are real) admit real 
solutions, but these solutions appear to be unstable.

\subsection{Addition of the Proctor term}
\label{subsec:snaking}

The branches of localised solutions shown in Figure~\ref{fig:bifurcation_diag}
(without the Proctor term) do not close. We now investigate the addition of the
Proctor term into the model, considering~\eqref{eq:PDE_nonlinear_Proctor} with
a rather large value of $\mu_{p}=-0.65$, chosen as to make the effects of this
term more pronounced. We also fix $f_{1}=0.2814$ and $f_{2}=-0.0721$ so that
the marginal stability curve has two minima at different heights.

Seeking localised solutions, we plotted (as before) $\mathcal{H}$ and
$\mathcal{F}$ (modified to include the Proctor term), looking for zero contours
of $\mathcal{H}(k_{+})-\mathcal{H}(k_{-})$ and
$\mathcal{F}(k_{+})-\mathcal{F}(k_{-})$, but we found no intersection of
contours for the two minima at different heights for this choice of parameters.
Notwithstanding this, we returned to wavenumbers $k_{-}=58/64$ and
$k_{+}=70/64$ and used a localised solution constructed from these constituent
wavenumbers as a starting point for timestepping and continuation. Part of the
resultant localised solution branch is shown in
Figure~\ref{fig:bifurcation_diag_Proctor}. We see that including the Proctor
term allows localised solutions of different widths on the same branch, rather
than lying on distinct branches as in Figure~\ref{fig:bifurcation_diag}. The
existence of the localised solutions is also limited to a finite range of $\mu$
values; we expect the upper limit is introduced owing to the band of Eckhaus
stable wavenumbers closing.
   
Figure~\ref{fig:solutions_Proctor} shows solutions at the saddle-nodes
indicated in Figure~\ref{fig:bifurcation_diag_Proctor}. We notice that at each
of these saddle-nodes the proportion of each pattern in the domain varies.
Lower down the branch the pattern with the smaller wavenumber fills more of the
domain, and conversely higher up the branch the  pattern with the larger
wavenumber fills more of the domain. This behaviour is qualitatively similar to
the snaking behaviour of localised solutions in the subcritical
Swift--Hohenberg equation \citep{Burke2006}, in that moving up the snaking
branch adds to the width of the spatially periodic part of the localised
solution -- but the details, with saddle-node bifurcations appearing at many
different places along the branch, are considerably more complicated, typical
of snaking in more complicated situations such as hexagons \citep{Lloyd2008} or
quasipatterns \citep{Subramanian2018} in two dimensions.
 \begin{figure}
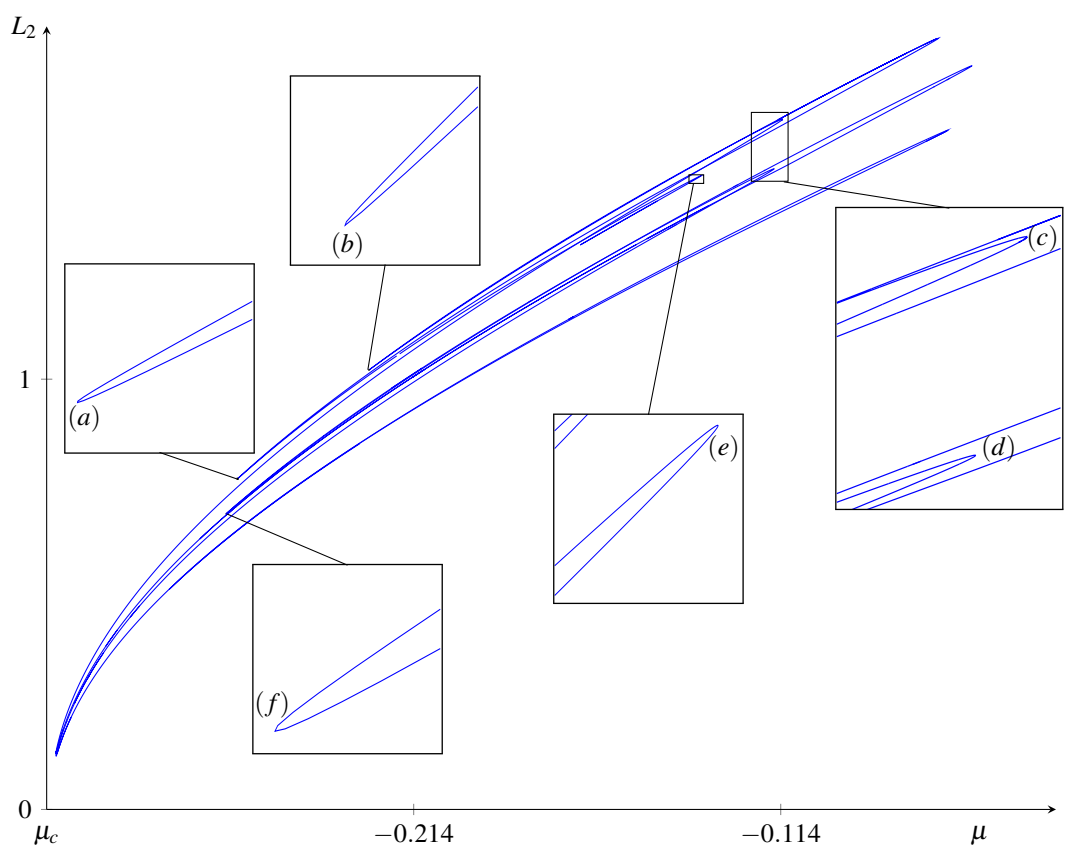

 \begin{center}
 \mbox{\beginpgfgraphicnamed{paper_figure_10}%
\begin{tikzpicture}
\begin{axis}[xtick={0.1,0.2},ytick={0,1},xticklabels={$-0.214$,$-0.114$},axis x line=bottom,axis y line=left,width=15cm,height=12cm,xmin=-0.0,xmax=0.275,ymin=0,ymax=1.82]
 \addplot[mark=none,color=blue] table[x index=0,y index=1]{ch6_p1.dat};
 \addplot[mark=none,color=blue] table[x index=0,y index=1]{ch6_p2.dat};
 \addplot[mark=none,color=blue] table[x index=0,y index=1]{ch6_proc3.dat};
 \addplot[mark=none,color=blue] table[x index=0,y index=1]{ch6_proc4.dat};
 \addplot[mark=none,color=blue] table[x index=0,y index=1]{ch6_p5.dat};
 \addplot[mark=none,color=blue] table[x index=0,y index=1]{ch6_proc6.dat};
 \addplot[mark=none,color=blue] table[x index=0,y index=1]{ch6_p7.dat};
 \addplot[mark=none,color=blue] table[x index=0,y index=1]{ch6_proc8.dat};
 \addplot[mark=none,color=blue] table[x index=0,y index=1]{ch6_proc9.dat};
 \addplot[mark=none,color=blue] table[x index=0,y index=1]{ch6_proc10.dat};
 \addplot[mark=none,color=blue] table[x index=0,y index=1]{ch6_proc11.dat};
 \addplot[mark=none,color=blue] table[x index=0,y index=1]{ch6_proc12.dat};
 \addplot[mark=none,color=blue] table[x index=0,y index=1]{ch6_proc13.dat};
 \addplot[mark=none,color=blue] table[x index=0,y index=1]{ch6_proc14.dat};
 \addplot[mark=none,color=blue] table[x index=0,y index=1]{ch6_p15.dat};
 \addplot[mark=none,color=blue] table[x index=0,y index=1]{ch6_proc16.dat};
 \addplot[mark=none,color=blue] table[x index=0,y index=1]{ch6_proc17.dat};
 \addplot[mark=none,color=blue] table[x index=0,y index=1]{ch6_p18.dat};
 \addplot[mark=none,color=blue] table[x index=0,y index=1]{ch6_p19.dat};
 \addplot[mark=none,color=blue] table[x index=0,y index=1]{ch6_proc20.dat};
 \addplot[mark=none,color=blue] table[x index=0,y index=1]{ch6_proc21.dat};
 \addplot[mark=none,color=blue] table[x index=0,y index=1]{ch6_p22.dat};
 \addplot[mark=none,color=blue] table[x index=0,y index=1]{ch6_p23.dat};
  \draw[-] (axis cs:0.192,1.46) -- (axis cs:0.202,1.46) -- (axis cs:0.202,1.62) -- (axis cs:0.192,1.62) -- cycle;
  \draw[-] (axis cs:0.175,1.455) -- (axis cs:0.179,1.455) -- (axis cs:0.179,1.475) -- (axis cs:0.175,1.475) -- cycle;
  \draw[-] (axis cs:0.08775,1.021) -- (axis cs:0.0879,1.021) -- (axis cs:0.0879,1.022) -- (axis cs:0.08775,1.022) -- cycle;
  \draw[-] (axis cs:0.0519,0.767) -- (axis cs:0.0521,0.767) -- (axis cs:0.0521,0.77) -- (axis cs:0.0519,0.77) -- cycle;
  \draw[-] (axis cs:0.048878,0.6868) -- (axis cs:0.048894,0.6868) -- (axis cs:0.048894,0.687) -- (axis cs:0.048878,0.687) -- cycle;


\end{axis}


\setlength{\fboxsep}{0pt}%
\setlength{\fboxrule}{0.5pt}%
\node [inner sep=0pt,outer sep=0pt] at (12,6) {\fbox{\includegraphics[trim=5mm 5mm 5mm 
5mm, clip, width=3cm, height=4cm]{ch6_procsn1}}};
\node [inner sep=0pt,outer sep=0pt] at (8,4) {\fbox{\includegraphics[trim=5mm 5mm 5mm 
5mm, clip, width=2.5cm, height=2.5cm]{ch6_procsn2}}};
\node [inner sep=0pt,outer sep=0pt] at (4.5,8.5) {\fbox{\includegraphics[trim=5mm 5mm 5mm 
5mm, clip, width=2.5cm, height=2.5cm]{ch6_procsn3}}};
\node [inner sep=0pt,outer sep=0pt] at (1.5,6) {\fbox{\includegraphics[trim=5mm 5mm 5mm 
5mm, clip, width=2.5cm, height=2.5cm]{ch6_procsn4}}};
\node [inner sep=0pt,outer sep=0pt] at (4,2) {\fbox{\includegraphics[trim=5mm 5mm 5mm 
5mm, clip, width=2.5cm, height=2.5cm]{ch6_procsn5}}};

 \draw (0.5,5.5) node[below] {$(a)$};
 \draw (4,7.8) node[below] {$(b)$};
 \draw (12.9,7.6) node[right] {$(c)$};
 \draw (12.3,4.8) node[right] {$(d)$};
 \draw (9,5.1) node[below] {$(e)$};
 \draw (3,1.1) node[above] {$(f)$};
 
 \draw[-] (1.5,4.75) --(2.55,4.39);    
 \draw[-] (4.5,7.25) --(4.27,5.85);    
 \draw[-] (12,8) --    (9.8,8.35);     
 \draw[-] (8,5.25) --  (8.6,8.33);     
 \draw[-] (4,3.25) --  (2.4,3.93);     

\draw (0,10.4) node[left] {$L_{2}$};
\draw (12.4,-0.1) node[below] {$\mu$};
\draw (0,-0.1) node[below] {$\mu_{c}$};
 \end{tikzpicture}
 \endpgfgraphicnamed}%
 \end{center}
 \caption{\label{fig:bifurcation_diag_Proctor} Branch of localised solutions
 with $\mu_{p}=-0.65$, $f_{1}=0.2814$, $f_{2}=-0.0721$. The critical value
 of the bifurcation parameter $\mu$ is $\mu_{c}=-0.314$. The interior 
 saddle-nodes are magnified and the labels $(a)-(e)$ correspond to
 the solution profiles shown in Figure~\ref{fig:solutions_Proctor}.}
 \end{figure}
 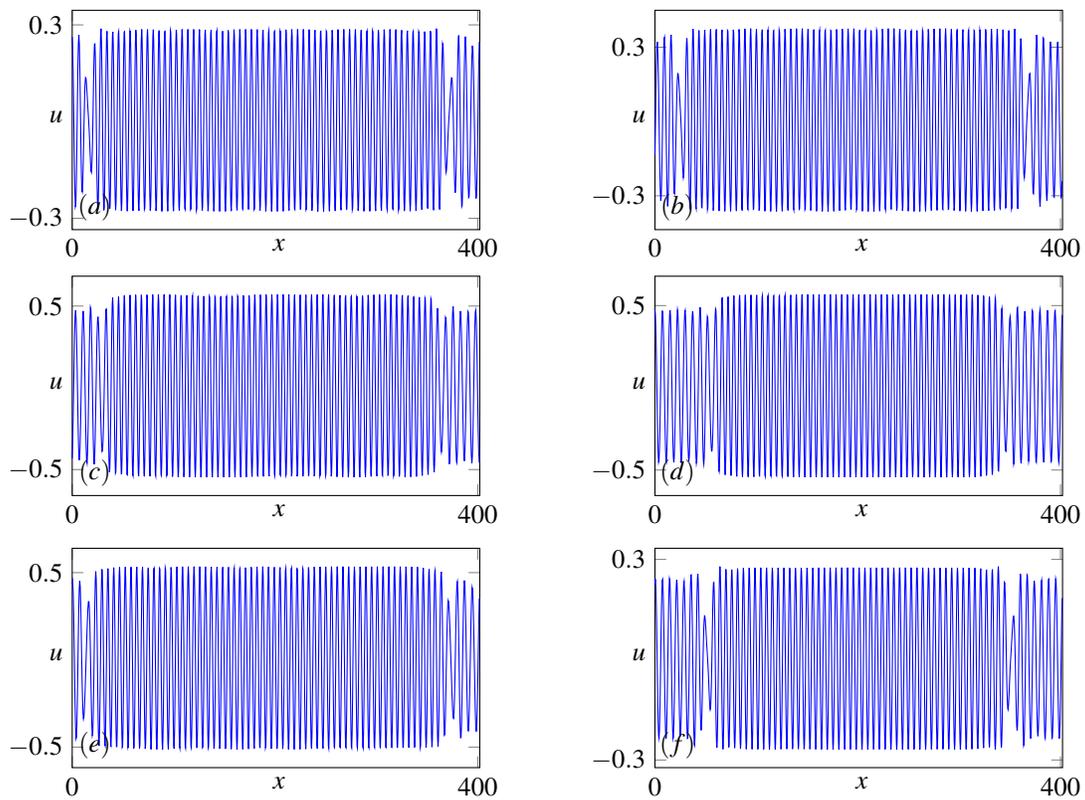
\begin{figure}
 \begin{center}
 \mbox{\beginpgfgraphicnamed{paper_figure_11}%
\begin{tabular}{cc}
\begin{tikzpicture}
\begin{axis}[xtick={0,0.995},xticklabels={0,400},ytick={-0.3,0.3},width=7cm,height=4.5cm,xmin=0,xmax=1]
\addplot[mark=none,color=blue] table[x index=0, y index=1]{ch6_p_msna.dat}; 
\end{axis}
\draw (0,1.5) node[left] {\hspace{1em} $u$};
\draw (2.75,0) node[below] {$x$};
\draw (0.3,0) node[above] {$(a)$};
\end{tikzpicture}
&
  \begin{tikzpicture}
\begin{axis}[xtick={0,0.995},xticklabels={0,400},ytick={-0.3,0.3},width=7cm,height=4.5cm,xmin=0,xmax=1]
\addplot[mark=none,color=blue] table[x index=0, y index=1]{ch6_p_msnb.dat}; 
\end{axis}
\draw (0,1.5) node[left] {\hspace{1em} $u$};
\draw (2.75,0) node[below] {$x$};
\draw (0.3,0) node[above] {$(b)$};
\end{tikzpicture}
\\
\begin{tikzpicture}
\begin{axis}[xtick={0,0.995},xticklabels={0,400},ytick={-0.5,0.5},width=7cm,height=4.5cm,xmin=0,xmax=1]
\addplot[mark=none,color=blue] table[x index=0, y index=1]{ch6_p_msnc.dat}; 
\end{axis}
\draw (0,1.5) node[left] {$u$};
\draw (2.75,0) node[below] {$x$};
\draw (0.3,0) node[above] {$(c)$};
  \end{tikzpicture}
&
  \begin{tikzpicture}
\begin{axis}[xtick={0,0.995},xticklabels={0,400},ytick={-0.5,0.5},width=7cm,height=4.5cm,xmin=0,xmax=1]
\addplot[mark=none,color=blue] table[x index=0, y index=1]{ch6_p_msnd.dat};
\end{axis}
\draw (0,1.5) node[left] {$u$};
\draw (2.75,0) node[below] {$x$};
\draw (0.3,0) node[above] {$(d)$};
\end{tikzpicture}
\\
\begin{tikzpicture}
\begin{axis}[xtick={0,0.995},xticklabels={0,400},ytick={-0.5,0.5},width=7cm,height=4.5cm,xmin=0,xmax=1]
\addplot[mark=none,color=blue] table[x index=0, y index=1]{ch6_p_msne.dat};
\end{axis}
\draw (0,1.5) node[left] {$u$};
\draw (2.75,0) node[below] {$x$};
\draw (0.3,0) node[above] {$(e)$};
  \end{tikzpicture}
&
\begin{tikzpicture}
\begin{axis}[xtick={0,0.995},xticklabels={0,400},ytick={-0.3,0.3},width=7cm,height=4.5cm,xmin=0,xmax=1]
\addplot[mark=none,color=blue] table[x index=0, y index=1]{ch6_p_msnf.dat};
\end{axis}
\draw (0,1.5) node[left] {$u$};
\draw (2.75,0) node[below] {$x$};
\draw (0.3,0) node[above] {$(f)$};
  \end{tikzpicture}
  \end{tabular}
 \endpgfgraphicnamed}%
 \end{center}
 \caption{\label{fig:solutions_Proctor} Solutions at the saddle-nodes
 indicated in Figure~\ref{fig:bifurcation_diag_Proctor}.}
 \end{figure}



\section{Discussion and conclusions}
\label{sec:conclusion}

The aim of this paper was to develop a new model
equation~\eqref{eq:PDE_nonlinear_Proctor} that captures qualitatively the
behaviour of pattern-forming problems with a quartic marginal stability curve,
and then explore the existence (and snaking) of localised solutions within this
model equation. There has been much progress in recent years developing a
framework for the understanding of localised solutions in the Swift--Hohenberg
equation \citep{Dawes2010,Knobloch2015}, where the existence of localised
patterns is interpreted in terms of a stable pattern existing at the same
parameter values as the stable trivial solution. Here we have shown that this
scenario holds in the case of the unfolding of a quartic minimum, where there
are coexisting stable patterns with two similar wavenumbers, and localised
solutions consisting of combinations of these. This work fits in with other
recent efforts that use Swift--Hohenberg-based models to explore problems with
localisation and snaking with multiple coexisting patterns
\citep{Knobloch2019,Alriheili2020,Subramanian2020}.

We computed the weakly nonlinear amplitude equation for the
model~\eqref{eq:PDE_nonlinear} and recovered a generalised Ginzburg--Landau
equation~\eqref{eq:generalised_GLE}. This allowed us to compute small-amplitude
solutions with different wavelengths and to classify the Eckhaus stable
patterns. Our work generalises that of \citet{Raitt1995} and \citet{Gelens2010}
(and corrects that of~\citet{Kozyreff2009}), to the case where the heights of
the minima in the marginal stability curve are different. We made use of a
first integral and a Lyapunov function to indentify candidate initial
conditions for finding localised solutions. Once the Proctor term, which allows
the wavenumber of maximum growth rate to depend on the bifurcation parameter,
is included in~\eqref{eq:PDE_nonlinear_Proctor}, the branches of localised
solutions join up, but the snaking we find is considerably more complicated
that the standard Swift--Hohenberg scenario.

An alternative approach to looking for localised solutions is to use spatial
dynamics, seeking only steady solutions of the model equation. For the
Swift--Hohenberg equation, this entails performing a normal form analysis of
the Hamiltonian--Hopf bifurcation, which describes the bifurcation from the
basic state, and contains the familiar homoclinic snaking of localised
solutions \citep{Woods1999}. In this framework, the existence of localised
solutions is determined by means of a geometric argument, whereby two integrals
of the normal form define a space which can be divided into regions which allow
or preclude the existence of localised solutions. In
Appendix~\ref{sec:appendix}, we derive a normal form for the bifurcation
occurring at the quartic minimum of the model equation, following the
derivation of the normal form for the Hamiltonian--Hopf bifurcation
\citep{Iooss1998a,Woods1999}. We use two different methods (following
\citet{Iooss1998a} and \citet{Burke2007a}) to find the coefficients in the
normal form. The two methods do not produce the same values for the
coefficients in the normal form, and in either case, there were terms appearing
in the normal form that one would have expected to be of higher order. As a
consequence of these terms we were unable to find normal form integrals,
including those one might expect by extension from the Hamiltonian--Hopf
analysis. This means we can not employ a similar geometric analysis to find
localised solutions. This normal form analysis merits further investigation. In
particular, it would be interesting to ascertain the reasons for the
difference between the normal form coefficients calculated by the two different
methods, and also to determine whether any variant of the normal form is 
integrable; the variant we have derived does not seem to be.

Having found numerical localised solutions in the model equation, we could
ask next whether examples more closely connected to reality might also have
them, for example the magnetised Taylor--Couette system \citep{Stefani2009} and
rotating magnetoconvection \citep{Cox2001, Chandrasekhar1961}, where there are
suggestions that marginal stability curves can change from having one to having
two minima. This last example offers the interesting possibility of exploring
two-dimensional patterns, potentially with regions of small hexagons embedded
in a background of large hexagons, for example. This would be a natural
extension to the study of localised solutions in the two-dimensional
Swift--Hohenberg equation \citep{Lloyd2008,Subramanian2018}.

The coalescence of the two minima is in itself an interesting problem. In one
dimensional systems with marginal stability curves with two minima far apart,
the natural approach is to reduce the problem to two coupled second-order
Ginzburg--Landau equations \citep{Dawes2008a}, and there are model equations
based on the Swift--Hohenberg equation that allow pattern formation on two
length scales \citep{Muller1994,Lifshitz1997,Rucklidge2012}. It would thus be
interesting to investigate how one would transition from the situation of two
well separated length scales to the unfolding of a quartic minimum.


\section*{Acknowledgements}

Thomas Wagenknecht (1974--2012) co-supervised much of the work presented in
this paper: DCB and AMR gratefully acknowledge his perceptive insight and
infectious enthusiasm. This project was inspired by conversations with Rainer
Hollerbach. We also acknowledge useful discussions and correspondence with
Jonathan Dawes, G\'erard Iooss, Edgar Knobloch, Gregory Kozyreff, Eric
Lombardi, Michael Proctor and Priya Subramanian. This research was supported by
a PhD studentship from the Science and Technology Facilities Council (STFC).


\appendix

\section{A normal form for the model equation}
\label{sec:appendix}                                

In this Appendix we derive the normal form for the model equation, with the
intention of extending the analysis of the Hamiltonian--Hopf bifurcation
in the Swift--Hohenberg equation \citep{Iooss1993,Woods1999} to this case of 
quadruply degenerate eigenvalues. Our approach is based on that of these 
authors (see also \citet{Haragus2011}), along with the work of 
\citet{Burke2007a}. For additional details of the calculations in this
Appendix, see \citet{Bentley2012}.

In general, we consider  dynamical systems of the form
\begin{equation} \label{eq:app:starting_point}
 \frac{d \mathbf{z}}{d x} = F (\mathbf{z}, \boldsymbol\varrho),
 \qquad
 \mathbf{z} \in \mathbb{R}^{n}, 
 \qquad
 \boldsymbol\varrho \in \mathbb{R}^{m},
\end{equation}
where we assume that
 \[
 F(0,0) = 0,
 \]
and also that there is a symmetry $\mathcal{R}$ such that
 \[
 F (\mathcal{R} \mathbf{z}, \boldsymbol\varrho) = - \mathcal{R} F(\mathbf{z}, \boldsymbol\varrho).
 \]
This symmetry $\mathcal{R}$ is known as a reversibility symmetry. We consider
reversible systems because of the invariance of the model equation
\eqref{eq:PDE_nonlinear} under spatial reversibility $x \rightarrow -x$. The 
parameter vector~$\boldsymbol\varrho=0$ is chosen so that the eigenvalues of 
the $\mathbf{z}=0$ equilibrium are all on the imaginary axis.

We aim to derive the normal form of the model equation near the bifurcation
This derivation essentially entails finding a
near-identity transform
 \[
 \mathbf{z} \rightarrow \tilde{\mathbf{z}} + \boldsymbol\Phi(\tilde{\mathbf{z}}, \boldsymbol\varrho),
 \]
such that we may write~\eqref{eq:app:starting_point} in the form
 \begin{equation}
 \label{eq:app:ending_point}
 \frac{d \tilde{\mathbf{z}}}{d x} = L_{0} \tilde{\mathbf{z}} + \mathbf{P}(\tilde{\mathbf{z}}, \boldsymbol\varrho) + o((|| \tilde{\mathbf{z}} || + || \boldsymbol\varrho ||)^{k_{p}}).
 \end{equation}
 Here $\boldsymbol\Phi(\tilde{\mathbf{z}}, \boldsymbol\varrho)$ and
$\mathbf{P}(\tilde{\mathbf{z}}, \boldsymbol\varrho)$ are ($n$-dimensional
vectors of) polynomials of degree $ \leq k_{p}$, and $L_{0}$ is a constant
coefficient matrix in Jordan normal form. The polynomial $\mathbf{P}$ satisfies
the so-called homological equation
 \begin{equation*}
 \mathbf{P} \left( e^{x L_{0}^{*}} \tilde{\mathbf{z}},\boldsymbol\varrho \right) 
  =
 e^{x L_{0}^{*}} \mathbf{P} \left( \tilde{\mathbf{z}}, \boldsymbol\varrho \right)
 \qquad \forall\, \tilde{\mathbf{z}}, \boldsymbol\varrho \hspace{1ex} \textrm{and} \hspace{1ex} x,
 \end{equation*}
where $L_{0}^{*}$ is the adjoint (conjugate transpose) of $L_{0}$. An
equivalent statement is found by
differentiating the equation with respect to~$x$ and evaluating at $x=0$, which 
gives
 \begin{equation*}
 \tilde{D} \mathbf{P} \left( \tilde{\mathbf{z}}, \boldsymbol\varrho \right) L_{0}^{*} \tilde{\mathbf{z}}
 =
 L_{0}^{*} \mathbf{P} \left( \tilde{\mathbf{z}}, \boldsymbol\varrho \right)
 \qquad
 \forall\, \tilde{\mathbf{z}} \hspace{1ex} \textrm{and} \hspace{1ex} \boldsymbol\varrho,
 \end{equation*}
where $\tilde{D} \mathbf{P} \left( \tilde{\mathbf{z}}, \boldsymbol\varrho \right)$ is the Jacobian matrix of $\mathbf{P}$.

There is some freedom in determining the polynomial $\mathbf{P}$; the idea is
to choose $\mathbf{P}$ to be as simple as possible. Of course, this freedom
means that there is not a unique normal form. Rather, the choice of
$\mathbf{P}$ is known as the style of the normal form. The two main styles are
the inner-product style popularised by \citet{Elphick1987a} and the $sl(2)$
style popularised by Cushman and Sanders \citep{Murdock2003}. We will use the
inner-product style, since this style is used for the Hamiltonian--Hopf
bifurcation\citep{Iooss1998a}, which describes the bifurcation from the basic
state in the Swift--Hohenberg equation.

In what follows, we will derive the normal form at the codimension 3
point $\mu=f_{1} = f_{2} = 0$, which corresponds to $\boldsymbol\varrho = 0$
in~\eqref{eq:app:starting_point}. So, the homological equation we will actually
use is
 \begin{equation} \label{eq:app:homological}
 \tilde{D} \mathbf{P} \left( \tilde{\mathbf{z}} \right) L_{0}^{*} \tilde{\mathbf{z}}
 =
 L_{0}^{*} \mathbf{P} \left( \tilde{\mathbf{z}} \right)
 \qquad
 \forall\, \tilde{\mathbf{z}}.
 \end{equation}
The parameters $\mu$, $f_{1}$ and $f_{2}$ can be added in as unfoldings once the normal form has been found.

\subsection{Linear part of the normal form}
\label{sec:app:linear_part}   

In this section, we write the dynamical system describing steady solutions of
the model equation. We then determine the linear part of the coordinate
transformation $\mathbf{z} \rightarrow \tilde{\mathbf{z}}$. By considering
steady solutions of our model equation~\eqref{eq:PDE_nonlinear} with $\mu=f_{1} 
= f_{2} = 0$, \textit{i.e.},
solutions of
 \begin{equation}
 \label{eq:app:degenerate_model}
 0 = - \left( 1 + \frac{\partial^{2}}{\partial x^{2}} \right)^{4} u 
 + n_{2} u^{2} + n_{3} u^{3},
 \end{equation}
 we can convert~\eqref{eq:app:degenerate_model} into a system of eight first-order ODEs, such
that we have the appropriate form~\eqref{eq:app:starting_point}. To do this, we
introduce new variables
 \[
 z_{1} = u,     \qquad
 z_{2} = u_{x}, \qquad 
 z_{3} = u_{xx}, \qquad 
 \dots, \qquad
 z_{8} = u_{xxxxxxx},
 \]
and write
 \begin{equation}
 \label{eq:app:spatial_dynamics}
 \frac{d \mathbf{z}}{dx} = F (\mathbf{z}) = L_0 \mathbf{z} + N(\mathbf{z}),
 \end{equation}
 where $\mathbf{z} = (z_{1}, z_{2}, z_{3}, z_{4}, z_{5}, z_{6}, z_{7},
z_{8})^{T}$. The linear and nonlinear parts of~\eqref{eq:app:spatial_dynamics} are
given by
 \[
 L_0 = \begin{pmatrix}
 0 & 1 & 0 & 0 & 0 & 0 & 0 & 0 \\
 0 & 0 & 1 & 0 & 0 & 0 & 0 & 0 \\
 0 & 0 & 0 & 1 & 0 & 0 & 0 & 0 \\
 0 & 0 & 0 & 0 & 1 & 0 & 0 & 0 \\
 0 & 0 & 0 & 0 & 0 & 1 & 0 & 0 \\
 0 & 0 & 0 & 0 & 0 & 0 & 1 & 0 \\
 0 & 0 & 0 & 0 & 0 & 0 & 0 & 1 \\
-1 & 0 & -4 & 0 & -6 & 0 & -4 & 0 
 \end{pmatrix}
 \qquad\text{and}\qquad
 N(\mathbf{z}) = \begin{pmatrix}
                  0 \\
                  0 \\
                  0 \\
                  0 \\
                  0 \\
                  0 \\
                  0 \\
                  n_{2} z_{1}^{2} + n_{3} z_{1}^{3}
                 \end{pmatrix}.
\]
The reversibility $\mathcal{R}$ acting on the elements of $\mathbf{z}$ is
defined as $\mathcal{R} z_{i} = (-1)^{i-1} z_{i}$ for $i=1,\ldots,8$. 

The first step is to transform $L_0$ into Jordan normal form. The eigenvalues
are $\lambda_{\pm}=\pm{i}$ with algebraic multiplicity~$4$ and geometric
multiplicity~$1$, so each eigenvalue has one eigenvector and three generalised
eigenvectors. These are readily found:
 \begin{equation} \label{eqn7_evec}
 \boldsymbol\zeta_{0} = \begin{pmatrix} 1 \\ i \\ -1 \\ -i \\ 1 \\ i \\ -1 \\ -i \end{pmatrix},
 \qquad
 \boldsymbol\zeta_{1} = \begin{pmatrix} 0 \\ 1 \\ 2i \\ -3 \\ -4i \\ 5 \\ 6i \\ -7 \end{pmatrix},
 \qquad
 \boldsymbol\zeta_{2} = \begin{pmatrix} 0 \\ 0 \\ 1 \\ 3i \\ -6 \\ -10i \\ 15 \\ 21i \end{pmatrix},
 \qquad
 \boldsymbol\zeta_{3} = \begin{pmatrix} 0 \\ 0 \\ 0 \\ 1 \\ 4i \\ -10 \\ -20i \\ 35 \end{pmatrix},
 \end{equation}
with 
 $L\boldsymbol\zeta_{0} = \lambda_{+}\boldsymbol\zeta_{0}$
 and
 $ L \boldsymbol\zeta_{j} = \lambda_{+}\boldsymbol\zeta_{j} + \boldsymbol\zeta_{j-1}$,
 with $j=1,2,3$.
We now define the linear transformation
 \begin{equation}
 \label{eq:app:linear_transform}
 \mathbf{z} = \boldsymbol\zeta_{0} A + \boldsymbol\zeta_{1} B + \boldsymbol\zeta_{2} C + \boldsymbol\zeta_{3} D + \bar{\boldsymbol\zeta_{0}} \bar{A} + \bar{\boldsymbol\zeta_{1}} \bar{B} + \bar{\boldsymbol\zeta_{2}} \bar{C} + \bar{\boldsymbol\zeta_{3}} \bar{D},
 \end{equation}
 where the overbar denotes complex conjugation, and $A$, $B$, $C$ and $D$ are
complex functions of~$x$. The transformed linear normal
form is $\frac{d \tilde{\mathbf{z}}}{d x}=L_{0} \tilde{\mathbf{z}}$, where
the transformed~$L_0$ and its adjoint~$L_{0}^{*}$ are
 \begin{equation*}
 L_{0} = \begin{pmatrix}
          i & 1 & 0 & 0 & 0 & 0 & 0 & 0 \\
          0 & i & 1 & 0 & 0 & 0 & 0 & 0 \\
          0 & 0 & i & 1 & 0 & 0 & 0 & 0 \\
          0 & 0 & 0 & i & 0 & 0 & 0 & 0 \\
          0 & 0 & 0 & 0 & -i & 1 & 0 & 0 \\
          0 & 0 & 0 & 0 & 0 & -i & 1 & 0 \\
          0 & 0 & 0 & 0 & 0 & 0 & -i & 1 \\
          0 & 0 & 0 & 0 & 0 & 0 & 0 & -i \\ 
         \end{pmatrix}
 \quad\text{and}\quad
 L_{0}^{*} = \begin{pmatrix}
          -i & 0 & 0 & 0 & 0 & 0 & 0 & 0 \\
          1 & -i & 0 & 0 & 0 & 0 & 0 & 0 \\
          0 & 1 & -i & 0 & 0 & 0 & 0 & 0 \\
          0 & 0 & 1 & -i & 0 & 0 & 0 & 0 \\
          0 & 0 & 0 & 0 & i & 0 & 0 & 0 \\
          0 & 0 & 0 & 0 & 1 & i & 0 & 0 \\
          0 & 0 & 0 & 0 & 0 & 1 & i & 0 \\
          0 & 0 & 0 & 0 & 0 & 0 & 1 & i \\ 
         \end{pmatrix}. 
 \end{equation*}
Thus, the linear part of the normal form is
 \begin{equation} \label{eq:app:nf_linear}
 \frac{d A}{d x} = i A + B,
 \qquad
 \frac{d B}{d x} = i B + C,
 \qquad
 \frac{d C}{d x} = i C + D,
 \qquad
 \frac{d D}{d x} = i D.
 \end{equation}
along with the complex conjugates of these.

\subsection{Nonlinear part of the normal form}
\label{sec:app:nonlinear_part}   

To determine the nonlinear part of the normal form
$\mathbf{P}(\tilde{\mathbf{z}})$ we make use of the homological
equation~\eqref{eq:app:homological}. We truncate at cubic order, setting
$k_{p}=3$ in~\eqref{eq:app:ending_point}. We also take in to account the linear
transformation in $\S$\ref{sec:app:linear_part}, now thinking of
$\tilde{\mathbf{z}}$ as $(A,B,C,D,\bar{A},\bar{B},\bar{C},\bar{D})$.

One possible approach to determining the nonlinear part of the normal form is
to suppose $\mathbf{P}(\tilde{\mathbf{z}})$ contains all possible quadratic and
cubic combinations of the components of $\tilde{\mathbf{z}}$, \textit{i.e.},
 \[
 \mathbf{P}(\tilde{\mathbf{z}}) = \sum_{i,j,k,l,m,n,o,p=0}^{3} \boldsymbol\Gamma_{ijklmnop} A^{i} B^{j} C^{k} D^{l} \bar{A}^{m} \bar{B}^{n} \bar{C}^{o} \bar{D}^{p},
 \]
such that $i+j+k+l+m+n+o+p=2$ or $3$. Then, plugging this into the homological
equation~\eqref{eq:app:homological} gives the terms and the combinations in
which they must appear.
This allows us to write the normal form as:
 \begin{subequations} \label{eq:app:normal_form_version_1}
 \begin{align} 
  \frac{d A}{d x} = iA + B  &+ \gamma_{1} |A|^{2} A + \gamma_{2} A \left( A \bar{B} - \bar{A} B \right) + \gamma_{3} A \left( A \bar{C} - |B|^{2} + \bar{A} C \right) \\ \nonumber
  & {}+ \gamma_{4} A \left( A \bar{D} - B \bar{C} + \bar{B} C - \bar{A} D \right) + \gamma_{5} \bar{A} \left( B^{2} - 2 A C \right) \\ \nonumber
  & {}+ 3 \gamma_{6} \left( \bar{B} \left( B^{2} - 2 A C \right) + \bar{A} \left( 3 A D - B C \right) \right),
 \end{align}
 \begin{align} 
   \frac{d B}{d x} = iB + C  &+ \gamma_{1} |A|^{2} B + \gamma_{2} B \left( A \bar{B} - \bar{A} B \right) + \gamma_{3} B \left( A \bar{C} - |B|^{2} + \bar{A} C \right) \\ \nonumber
  & {}+ \gamma_{4} B \left( A \bar{D} - B \bar{C} + \bar{B} C - \bar{A} D \right) + \gamma_{5} \bar{B} \left( B^{2} - 2 A C \right) \\ \nonumber
  & {}+ \gamma_{6} \left(2 \bar{A} \left( 3 B D - 2 C^{2} \right)  + \bar{B} \left( 3 A D - B C \right) + 4 \bar{C} \left( B^{2} - 2 A C \right) \right) \\ \nonumber
  & {}+ \gamma_{7} |A|^{2} A + \gamma_{8} A \left( A \bar{B} - \bar{A} B \right) + \gamma_{9} A \left( A \bar{C} - |B|^{2} + \bar{A} C \right) \\ \nonumber
  & {}+ \gamma_{10} A \left( A \bar{D} - B \bar{C} + \bar{B} C - \bar{A} D \right) + 2 \gamma_{11} \bar{A} \left( B^{2} - 2 A C \right) \\ \nonumber
  & {}+ \gamma_{12} \left( \bar{B} \left( B^{2} - 2 A C \right) + \bar{A} \left( 3 A D - B C \right) \right),
 \end{align}
 \begin{align} 
   \frac{d C}{d x} = iC + D  &+ \gamma_{1} |A|^{2} C + \gamma_{2} C \left( A \bar{B} - \bar{A} B \right) + \gamma_{3} C \left( A \bar{C} - |B|^{2} + \bar{A} C \right) \\ \nonumber
  & {}+ \gamma_{4} C \left( A \bar{D} - B \bar{C} + \bar{B} C - \bar{A} D \right)  + \gamma_{5} \bar{C} \left( B^{2} - 2 A C \right) \\ \nonumber
  & {}+ \gamma_{6} \left( 2 \bar{B} \left( 3 B D - 2 C^{2} \right) - \bar{C} \left( 3 A D - B C \right) + 3 \bar{D} \left( B^{2} - 2 A C \right) \right) \\ \nonumber
  & {}+ \gamma_{7} |A|^{2} B + \gamma_{8} B \left( A \bar{B} - \bar{A} B \right) + \gamma_{9} B \left( A \bar{C} - |B|^{2} + \bar{A} C \right) \\ \nonumber
  & {}+ \gamma_{10} B \left( A \bar{D} - B \bar{C} + \bar{B} C - \bar{A} D \right) + \gamma_{11} \left( \bar{B} \left( B^{2} -2 A C \right) - \bar{A} \left( 3 A D - B C \right) \right) \\ \nonumber
  & {}+ \gamma_{12} \left( 2 \bar{B} \left( 3 A D - B C \right) + 3 \bar{C} \left( B^{2} - 2 A C \right) - \bar{A} \left( 3 B D - 2 C^{2} \right) \right) + \gamma_{13} |A|^{2} A \\ \nonumber
  & {}+ \gamma_{14} A \left( A \bar{B} - \bar{A} B \right) + \gamma_{15} A \left( A \bar{C} - |B|^{2} + \bar{A} C \right) \\ \nonumber
  & {}+ \gamma_{16} A \left( A \bar{D} - B \bar{C} + \bar{B} C - \bar{A} D \right) + \gamma_{17} \bar{A} \left( B^{2} - 2 A C \right) + \gamma_{18} \left( \bar{A} \left( 3 B D - 2 C^{2} \right) \right. \\ \nonumber
  & \qquad {}- \left. \bar{B} \left( 3 A D - B C \right) - \bar{C} \left( B^{2} - 2 A C \right) \right) + \gamma_{19} \left( \bar{B} \left( B^{2} - 2 A C \right) + \bar{A} \left( 3 A D - B C \right) \right),
 \end{align}
 \begin{align} 
   \frac{d D}{d x} = iD &+ \gamma_{1} |A|^{2} D + \gamma_{2} D \left( A \bar{B} - \bar{A} B \right) + \gamma_{3} D \left( A \bar{C} - |B|^{2} + \bar{A} C \right) \\ \nonumber
  & {}+ \gamma_{4} D \left( A \bar{D} - B \bar{C} + \bar{B} C - \bar{A} D \right) + \gamma_{5} \bar{D} \left( B^{2} - 2 A C \right) \\ \nonumber
  & {}+ \gamma_{6} \left( 2 \bar{C} \left( 3 B D - 2 C^{2} \right) - 3 \bar{D} \left( 3 A D - B C \right) \right) + \gamma_{7} |A|^{2} C \\ \nonumber
  & {}+ \gamma_{8} C \left( A \bar{B} - \bar{A} B \right) + \gamma_{9} C \left( A \bar{C} - |B|^{2} + \bar{A} C \right) + \gamma_{10} C \left( A \bar{D} - B \bar{C} + \bar{B} C - \bar{A} D \right) \\ \nonumber
  & {}- \gamma_{11} \bar{B} \left( 3 A D - B C \right) + \gamma_{12} \left( 3 \bar{C} \left( B^{2} - 2 A C \right) + 6 \bar{D} \left( 3 A D - B C \right) \right. \\ \nonumber
  & \qquad {}- \left. \bar{B} \left( 3 B D - 2 C^{2} \right) \right) + \gamma_{13} |A|^{2} B + \gamma_{14} B \left( A \bar{B} - \bar{A} B \right) + \gamma_{15} B \left( A \bar{C} - |B|^{2} + \bar{A} C \right) \\ \nonumber
  & {}- \gamma_{17} \bar{A} \left( 3 A D - B C \right) + \gamma_{18} \left( \bar{B} \left( 3 B D - 2 C^{2} \right) - 2 \bar{C} \left( B^{2} - 2 A C \right) \right. \\ \nonumber
  & \qquad {}- \left. 3 \bar{D} \left( 3 A D - B C \right) \right) + \gamma_{19} \left( 2 \bar{A} \left( 3 B D - 2 C^{2} \right) - \bar{B} \left( 3 A D - B C \right) \right) \\ \nonumber
  & {}+ \gamma_{20} |A|^{2} A + \gamma_{21} A \left( A \bar{B} - \bar{A} B \right) + \gamma_{22} A \left( A \bar{C} - |B|^{2} + \bar{A} C \right) \\ \nonumber
  & {}+ \gamma_{23} A \left( A \bar{D} - B \bar{C} + \bar{B} C - \bar{A} D \right) + \gamma_{24} \bar{A} \left( B^{2} - 2 A C \right) + \gamma_{25} \left( \bar{B} \left( 3 A D - B C \right) \right. \\ \nonumber
  & \qquad {}- \left. \bar{A} \left( 3 B D - 2 C^{2} \right)  + \bar{C} \left( B^{2} - 2 A C \right) \right) + \gamma_{26} \left( \bar{A} \left( 3 A D - B C \right) \right. \\ \nonumber
  & \qquad {}+ \left. \bar{B} \left( B^{2} - 2 A C \right) \right),
 \end{align}
\end{subequations}
where the coefficients $\gamma_{i}$, $i=1,\ldots,26$ are to be determined by
transforming the nonlinear term $n_2u^2+n_3u^3$. One may notice
that~\eqref{eq:app:normal_form_version_1} contains no quadratic terms; the
requirement that we satisfy the homological equation~\eqref{eq:app:homological}
excludes them. Had we chosen a different normal form style, \textit{e.g.}, the
$sl(2)$ style \citep{Murdock2003}, then it is possible
that~\eqref{eq:app:normal_form_version_1} would have contained quadratic terms.

An alternative approach to constructing the normal form is to find first
integrals of the homological equation and construct the polynomials
$\mathbf{P}(\tilde{\mathbf{z}})$ using these, following \citet{Elphick1987a}.
For details, see \citet{Bentley2012}. The end result is a set of first
integrals $c_1$, \dots, $c_7$, given in Table~\ref{tab:integrals}. Note that
$c_2$, $c_4$ and $c_6$ contain $\log(A)$, so we also use $w_1$ and $w_2$, which
are combinations of the first seven with the $\log(A)$ dependence eliminated.
We note that the integrals $c_{1}$, $c_{2}$ and $c_{3}$ are also integrals of
the characteristic system in the four-dimensional Hamiltonian--Hopf case
\citep{Iooss1998a}.
 \begin{table}
\begin{center}
\begin{tabular}{|c|l|}
\hline
$c_{1}$ & {\vphantom{\Big(}} $A\bar{A}$ \\
 \hline
$c_{2}$ & {\vphantom{\Big(}} $\frac{iB}{A} + \log(A)$ \\
\hline
$c_{3}$ & {\vphantom{\Big(}} $i(A\bar{B}-\bar{A}B)$ \\
\hline
$c_{4}$ & {\vphantom{\Big(}} $\frac{C}{A} - i\frac{B}{A} \log(A) - \frac{1}{2} (\log(A))^{2}$ \\
\hline
$c_{5}$ & {\vphantom{\Big(}} $A\bar{C}-B\bar{B}+\bar{A}C$ \\
\hline
$c_{6}$ & {\vphantom{\Big(}} $- i\frac{D}{A} - \frac{C}{A} \log(A) + i\frac{B}{2 A} (\log(A))^{2} + \frac{1}{6} (\log(A))^{3}$ \\
\hline
$c_{7}$ & {\vphantom{\Big(}} $i(A\bar{D}-B\bar{C}+\bar{B}C-\bar{A}D)$ \\
\hline
$w_{1}$ & {\vphantom{\Big(}} $- \frac{1}{A^{2}} \left( B^{2} - 2 A C \right)$ \\
\hline
$w_{2}$ & {\vphantom{\Big(}} $ - \frac{i}{A^{2}} \left( \left( 3 A D - B C \right) + \frac{B}{A} \left( B^{2} - 2 A C \right) \right)$ \\
\hline
\end{tabular}
 \caption{The first integrals of the homological equation.}
 \label{tab:integrals}
\end{center}
\end{table}

From the first integrals of the homological equation, we can construct the
nonlinear part of the normal form. We make the change of variables
 \begin{equation*}
 (A,B,C,D,\bar{A},\bar{B},\bar{C},\bar{D}) \rightarrow (A,c_{1},c_{2},c_{3},c_{5},c_{7},w_{1},w_{2}),
 \end{equation*}
and solve the homological equation in these new variables
\citep[see][]{Bentley2012}, giving, for the first equation in the normal form:
 \[
 \frac{dA}{dx} = iA + B + P_{1}(A,B,C,D,\bar{A},\bar{B},\bar{C},\bar{D}) 
               = iA + B + A \varphi(c_{1},c_{2},c_{3},c_{5},c_{7},w_{1},w_{2}),
 \]
some some arbitrary function~$\varphi$, provided that $P_{1}=A\varphi$ is a
polynomial in its eight arguments. Here we use $w_1$ and $w_2$ in preference to
$c_4$ and~$c_6$.

In the derivation of the normal form for the Hamiltonian--Hopf bifurcation, the
equivalent equation at this stage is $P_{1}(A,B,\bar{A},\bar{B}) = A
\varphi(c_{1},c_{2},c_{3})$. The argument is then that $\varphi$ is a
polynomial in $c_{1}$ and $c_{3}$, and independent of $c_{2}$. This is because
of the log dependence of $c_{2}$: as $A\rightarrow0$, the logarithmic behaviour
of $c_{2}$ does not match the polynomial behaviour of $P_{1}$, and thus
$\varphi$ must be independent of $c_{2}$. This argument follows
through to our case as far as $c_{2}$ is concerned, but there are additional 
considerations regarding $w_{1}$ and $w_{2}$, which have $A^2$ in their 
denominators. This dependence in $A\varphi$ is eliminated by taking certain
combinations of $w_{1}$ and $w_{2}$: for example,
$Ac_{1}w_{1}=-\bar{A}(B^{2}-2AC)$, which is fine, as is
$A(c_{1}w_{2}+w_{1}c_{3})=\bar{A}(3AD-BC)+\bar{B}(B^{2}-2AC)$, while
$Ac_{1}w_{2}$ has an $A$ in the denominator and so is not a polynomial. In
fact, only the two combinations $c_{1}w_{1}$ and $c_{1}w_{2}+c_{3}w_{1}$ are
needed for $P_{1}$ for $\frac{dA}{dx}$, but additional combinations appear in
the other three equations.
 
After conputing these and (re)labelling the arbitrary functions as $P$, $Q$,
$R$ and~$S$, to be consistent with the notation of the normal form of the
Hamiltonian--Hopf bifurcation \citep{Burke2007a,Iooss1998a}, we have the
eight-dimensional normal form
 \begin{subequations}
 \label{eq:app:normal_form_version_2}
 \begin{equation} \label{eq:app:normal_form_version_2a}
 A_{x} = iA + B + iA P(c_{1}, c_{3}, c_{5}, c_{7}, w_{1}, w_{2}),
 \end{equation}
 \begin{equation} \label{eq:app:normal_form_version_2b}
 B_{x} = iB + C + iB P(c_{1}, c_{3}, c_{5}, c_{7}, w_{1}, w_{2}) + A Q(c_{1}, c_{3}, c_{5}, c_{7}, w_{1}, w_{2}),
 \end{equation}
 \begin{align} \nonumber
 C_{x} &= iC + D + iC P(c_{1}, c_{3}, c_{5}, c_{7}, w_{1}, w_{2}) + B Q(c_{1}, c_{3}, c_{5}, c_{7}, w_{1}, w_{2}) \\ \label{eq:app:normal_form_version_2c}
       &\qquad + iA R(c_{1}, c_{3}, c_{5}, c_{7}, w_{1}, w_{2}),
 \end{align}
 \begin{align} \nonumber
 D_{x} &= iD + iD P(c_{1}, c_{3}, c_{5}, c_{7}, w_{1}, w_{2}) + C Q(c_{1}, c_{3}, c_{5}, c_{7}, w_{1}, w_{2}) \\ \label{eq:app:normal_form_version_2d}
       &\qquad + iB R(c_{1}, c_{3}, c_{5}, c_{7}, w_{1}, w_{2}) + A S(c_{1}, c_{3}, c_{5}, c_{7}, w_{1}, w_{2}).
 \end{align}
 \end{subequations}
The functions $P$, $Q$, $R$ and~$S$ are understood to include only those
combinations of $w_1$ and $w_2$ that result in polynomial contributions to the
normal form. The nonlinear terms up to cubic order are:
 \[
 P(c_{1}, c_{3}, c_{5}, c_{7}, w_{1}, w_{2}) = P_{1} c_{1} + P_{2} c_{3} + P_{3} c_{5} + P_{4} c_{7} + 3T_{1} (c_{1} w_{2} + c_{3} w_{1}) - iT_{2} c_{1} w_{1},
 \]
 \begin{align*}
 Q(c_{1}, c_{3}, c_{5}, c_{7}, w_{1}, w_{2}) &= Q_{1} c_{1} + Q_{2} c_{3} + Q_{3} c_{5} + Q_{4} c_{7} + T_{1} (c_{3} w_{2} - 4 c_{5} w_{1} + c_{1} w_{1}^{2}) \\  
  &\qquad - i T_{2} c_{3} w_{1} - T_{3} c_{1} w_{1} - i T_{4}(c_{3} w_{1} + c_{1} w_{2}),
 \end{align*}
 \begin{align*} 
 R(c_{1}, c_{3}, c_{5}, c_{7}, w_{1}, w_{2}) &= R_{1} c_{1} + R_{2} c_{3} + R_{3} c_{5} + R_{4} c_{7} \\ 
  &\qquad + T_{1} (3c_{7} w_{1} - c_{5} w_{2} - 2 c_{1} w_{1} w_{2} - 2 c_{3} w_{1}^{2}) - i T_{2} (c_{5} w_{1} - c_{1} w_{1}^{2}) \\ 
   &\qquad - \tfrac{1}{2} T_{3} (c_{1} w_{2} - c_{3} w_{1}) + i T_{4} (2 c_{3} w_{2} - 3 c_{5} w_{1} + 2 c_{1} w_{1}^{2}) \\
    &\qquad + i T_{5} c_{1} w_{1} + i T_{6} (c_{3} w_{2} - c_{5} w_{1} + c_{1} w_{1}^{2}) - \tfrac{1}{2} T_{7} (c_{1} w_{2} + c_{3} w_{1}),
\end{align*}
\begin{align*}
 S(c_{1}, c_{3}, c_{5}, c_{7}, w_{1}, w_{2}) &= S_{1} c_{1} + S_{2} c_{3} + S_{3} c_{5} + S_{4} c_{7} \\ 
  &\qquad + T_{1} (2 c_{1} w_{2}^{2} - 3 c_{7} w_{2} + 2 c_{3} w_{1} w_{2} + c_{5} w_{1}^{2}) - i T_{2}(c_{7} w_{1} - c_{3} w_{1}^{2}) \\ 
   &\qquad - \tfrac{1}{2} T_{3} (c_{3} w_{2} - c_{1} w_{1}^{2}) + i T_{4} (4 c_{1} w_{1} w_{2} + 4 c_{3} w_{1}^{2} - 3 c_{5} w_{2} - 6 c_{7} w_{1}) \\ 
    &\qquad - i T_{5} c_{1} w_{2} + i T_{6} (2 c_{1} w_{1} w_{2} + c_{3} w_{1}^{2} - 2 c_{5} w_{2} + 3 c_{7} w_{1}) \\ 
    &\qquad + \tfrac{1}{2} T_{7} (c_{3} w_{2} + c_{1} w_{1}^{2}) - T_{8} c_{1} w_{1} + T_{9} (c_{3} w_{2} - c_{5} w_{1} + c_{1} w_{1}^{2}) \\ 
    &\qquad + i T_{10} (c_{3} w_{1} + c_{1} w_{2}), 
\end{align*}
where the $P_{i}$, $Q_{i}$, $R_{i}$, $S_{i}$ and $T_{j}$, $i=1,\ldots,4$,
$j=1,\ldots,10$ are real coefficients (with the prefactor~$i$ included in the 
cases of $T_2$, $T_4$, $T_5$, $T_6$ and~$T_{10}$). These terms may seem
somewhat arbitrary, but are in fact very specific combinations to match the
terms found by solving the homological equation (compare
with~\eqref{eq:app:normal_form_version_1}). The relation between the normal
form coefficients here and the normal form coefficients
in~\eqref{eq:app:normal_form_version_1} is given in
Table~\ref{tab:coefficients}.
 \begin{table}
 \begin{tabular}{|c|c|c|c|c|c|}
\hline
 {\vphantom{\Big(}} $\gamma_{1} = i P_{1}$ & $\gamma_{2}=-P_{2}$ & $\gamma_{3} = i P_{3}$ & $\gamma_{4}=-P_{4}$ & $\gamma_{5} = -T_{2}$ & $\gamma_{6} = T_{1}$ \\
\hline
 {\vphantom{\Big(}} $\gamma_{7}=Q_{1}$ &  $\gamma_{8} = i Q_{2}$ & $\gamma_{9}=Q_{3}$ & $\gamma_{10} = i Q_{4}$ & $\gamma_{11}=T_{3}$ & $\gamma_{12} = -T_{4}$ \\
\hline
 {\vphantom{\Big(}} $\gamma_{13} = i R_{1}$ & $\gamma_{14}=-R_{2}$ &  $\gamma_{15} = i R_{3}$ & $\gamma_{16}=-R_{4}$ & $\gamma_{17} = T_{5}$ & $\gamma_{18}=2 T_{6}$ \\
\hline
 {\vphantom{\Big(}} $\gamma_{19} = -T_{7}/2$ & $\gamma_{20} = S_{1}$ & $\gamma_{22} = i S_{2}$ &  $\gamma_{22} =  S_{3}$ & $\gamma_{23}=i S_{4}$ & $\gamma_{24} = T_{8}$ \\
\hline
 {\vphantom{\Big(}} $\gamma_{25}=T_{9}$ & $\gamma_{26} = T_{10}$ & & & & \\
\hline
 \end{tabular}
\caption{\label{tab:coefficients} Relation between the normal coefficients in~\eqref{eq:app:normal_form_version_1} and the normal form coefficients in~\eqref{eq:app:normal_form_version_2}.}
\end{table}

Having found the terms present in the normal form, we now wish to find the
coefficients of these terms. In the following sections, we will describe two
methods to do this: by solving a linear system of equations derived from the
system of ODEs~\eqref{eq:app:spatial_dynamics}, following \citet{Iooss1998a},
and an asymptotic scaling method, following \citet{Burke2007a}.

\subsection{Determining the normal form coefficients I: 
            Nonlinear coordinate transform}

We have so far described the linear
transformation~\eqref{eq:app:linear_transform}. Following \citet{Iooss1998a},
we now add nonlinear terms to the transformation, in particular a polynomial
$\boldsymbol\Phi(\tilde{\mathbf{z}})$, such that we have
 \begin{equation}
 \label{eq:app:nonlinear_transform}
 \mathbf{z} = \boldsymbol\zeta_{0} A + \boldsymbol\zeta_{1} B + \boldsymbol\zeta_{2} C + \boldsymbol\zeta_{3} D + \bar{\boldsymbol\zeta_{0}} \bar{A} + \bar{\boldsymbol\zeta_{1}} \bar{B} + \bar{\boldsymbol\zeta_{2}} \bar{C} + \bar{\boldsymbol\zeta_{3}} \bar{D} + \boldsymbol\Phi(\mathbf{\tilde{z}}).
 \end{equation}
We fix $\boldsymbol\Phi$ such that it contains only quadratic and cubic terms.
Substituting this into~\eqref{eq:app:spatial_dynamics} and matching like powers 
of the variables results in a relationship between the parameters $n_2$ 
and~$n_3$ in the model~\eqref{eq:app:degenerate_model} and the normal form 
coefficients in~\eqref{eq:app:normal_form_version_2}:
 \begin{align} \label{eqn7_nf_coeffs1}
 P_{1} &= -\frac{1935}{5824} n_{3} - \frac{6686165}{6368544} n_{2}^{2} , \qquad P_{2} = -\frac{1418305}{1207224} n_{2}^{2}, \\ \nonumber
 P_{3} &= \frac{12194005}{6613488} n_{2}^{2} , \qquad P_{4} = -\frac{6220189}{8424324} n_{2}^{2}, \\ \nonumber
 Q_{1} &= -\frac{75}{224} n_{3} - \frac{61235}{81648} n_{2}^{2} , \qquad Q_{2} = -\frac{645}{5824} n_{3} - \frac{5539735}{6368544} n_{2}^{2}, \\ \nonumber
 Q_{3} &= \frac{2922859}{1810836} n_{2}^{2} , \qquad Q_{4} = \frac{4594355}{6613488} n_{2}^{2}, \\ \nonumber
 R_{1} &= \frac{9}{32} n_{3} + \frac{1483}{3888} n_{2}^{2} , \qquad R_{2} = \frac{25}{224} n_{3} + \frac{45649}{81648} n_{2}^{2}, \\ \nonumber
 R_{3} &= -\frac{135}{5824} n_{3} - \frac{513995}{909792} n_{2}^{2} , \qquad R_{4} = - \frac{923965}{1207224} n_{2}^{2}, \\ \nonumber
 S_{1} &= \frac{3}{16} n_{3} + \frac{163}{648} n_{2}^{2} , \qquad S_{2} = \frac{3}{32} n_{3} + \frac{473}{3888} n_{2}^{2}, \\ \nonumber
 S_{3} &= - \frac{5}{224} n_{3} - \frac{967}{3024} n_{2}^{2} , \qquad S_{4} = - \frac{15}{5824} n_{3} - \frac{918133}{6368544} n_{2}^{2}, \\ \nonumber
 T_{1} &= \frac{5400535}{2808108} n_{2}^{2} , \qquad T_{2} = \frac{12625255}{4408992} i n_{2}^{2}, \\ \nonumber
 T_{3} &= -\frac{2005877}{1810836} n_{2}^{2} , \qquad T_{4} = -\frac{528565}{944784} i n_{2}^{2}, \\ \nonumber
 T_{5} &= -\frac{225}{1456} i n_{3} -\frac{221885}{530712} i n_{2}^{2} , \qquad T_{6} = \frac{4026445}{6613488} i n_{2}^{2}, \\ \nonumber
 T_{7} &= \frac{1142417}{1810836} n_{2}^{2} , \qquad T_{8} = -\frac{5}{32} n_{3} -\frac{815}{3888} n_{2}^{2}, \\ \nonumber
 T_{9} &= -\frac{1718057}{1810836} n_{2}^{2}, \qquad T_{10} = -\frac{135}{2912} i n_{3} - \frac{286375}{1061424} i n_{2}^{2}. 
 \end{align}
All the normal form coefficients are either purely real or purely imaginary,
which is a consequence of the reversibility symmetry.

\subsection{Determining the normal form coefficients II: Asymptotic scaling method}

This method involves expanding both the steady model
equation~\eqref{eq:app:degenerate_model} and the normal form
equations~\eqref{eq:app:normal_form_version_2} in powers of a small
parameter~$\epsilon$, following \citet{Burke2007a}. The normal form
coefficients can then be found by matching the equations at each order of~$\epsilon$.

\subsubsection{Model equation expansion}
\label{sec:app:model_asymptotics}

We introduce a small parameter~$\epsilon$, and define a long length scale
$X=\epsilon^{1/2}x$, as we did in the weakly nonlinear analysis
in~$\S$\ref{sec:wnlt}. We then expand $u(x)$ in terms of this small parameter,
\textit{i.e.},
 \begin{equation} \label{eqn7_shs_exp}
 u(x) = \sum_{n=2}^{12} \epsilon^{n/2} u_{n}(x,X). 
 \end{equation}
The summation index runs from $n=2$ because the amplitude of $u$ is
$\mathcal{O}(\epsilon)$ (see $\S$\ref{sec:wnlt}). We need to go to $n=12$ in
order to determine all the coefficients in the normal
form~\eqref{eq:app:normal_form_version_2}.

Substituting~\eqref{eqn7_shs_exp} into~\eqref{eq:app:degenerate_model}, we obtain equations to be solved at each order of $\epsilon^{1/2}$, \textit{i.e.},
\begin{align*} 
 \mathcal{O} (\epsilon): 0 &= - \left( 1 + \partial_{xx} \right)^{4} u_{2} \equiv - \mathcal{L} u_{2}, \\ 
 \mathcal{O} (\epsilon^{3/2}): 0 &= - \mathcal{L} u_{3} - 8 \partial_{xX} \left( 1 + \partial_{x}^{2} \right)^{3} u_{2}, \\ 
 \mathcal{O} (\epsilon^{2}): 0 &= - \mathcal{L} u_{4} - 8 \partial_{xX} \left( 1 + \partial_{x}^{2} \right)^{3} u_{3} - 4 \partial_{X}^{2} \left( 1 + 9 \partial_{x}^{2} + 15 \partial_{x}^{4} + 7 \partial_{x}^{6} \right) u_{2} + n_{2} u_{2}^{2}, \\ 
 \mathcal{O} (\epsilon^{5/2}): 0 &= - \mathcal{L} u_{5} - 8 \partial_{xX} \left( 1 + \partial_{x}^{2} \right)^{3} u_{4} - 4 \partial_{X}^{2} \left( 1 + 9 \partial_{x}^{2} + 15 \partial_{x}^{4} + 7 \partial_{x}^{6} \right) u_{3} \\ 
 &\qquad {} - 8 \partial_{X}^{3} \partial_{X} \left( 3 + 10 \partial_{x}^{2} + 7 \partial_{x}^{4} \right) u_{2} + 2 n_{2} u_{2} u_{3}, \\ 
 \mathcal{O} (\epsilon^{3}): 0 &= - \mathcal{L} u_{6} - 8 \partial_{xX} \left( 1 + \partial_{x}^{2} \right)^{3} u_{5} - 4 \partial_{X}^{2} \left( 1 + 9 \partial_{x}^{2} + 15 \partial_{x}^{4} + 7 \partial_{x}^{6} \right) u_{4} \\ 
 &\qquad {} - 8 \partial_{X}^{3} \partial_{X} \left( 3 + 10 \partial_{x}^{2} + 7 \partial_{x}^{4} \right) u_{3} - 2 \partial_{X}^{4} \left( 3 + 30 \partial_{x}^{2} + 35 \partial_{x}^{4} \right) u_{2} + n_{2} \left( u_{3}^{2} + 2 u_{2} u_{4} \right) + n_{3} u_{2}^{3},
\end{align*}
and similarly for higher orders, up to $\mathcal{O}(\epsilon^{6})$. This
analysis is equivalent to the one performed in~$\S$\ref{sec:wnlt}, though there
is a factor of two difference in the subscript of $u$ between the two
calculations.

The leading order solution is given by
 \[
 u_{2}(x,X) = A_{2}(X) e^{ix} + c.c.,
 \]
and similarly from $\mathcal{O}(\epsilon^{3/2})$ we have
 \[
u_{3}(x,X) = A_{3}(X) e^{ix} + c.c.. 
 \]
In the weakly nonlinear analysis in $\S$\ref{sec:wnlt}, we set $u_{3/2}=0$ (equivalently $u_{3}=0$ here); in this analysis we keep $u_{3}$ non-zero. The difference is of no consequence, however. 

Proceeding to $\mathcal{O}(\epsilon^{2})$, the ansatz $u_{4} = \lambda_{4} + A_{4}(X) e^{ix} + B_{4}(X) e^{2ix} + c.c.$, where $\lambda_{4}$ is real, leads to the solution
 \[
 \lambda_{4} = 2 n_{2} |A_{2}|^{2}, \qquad B_{4} = \frac{n_{2}}{81} A_{2}^{2}, \qquad \bar{B}_{4} = \frac{n_{2}}{81} \bar{A}_{2}^{2},
 \]
and $A_{4}$, $\bar{A}_{4}$ are as yet unknown. We have dropped the explicit $X$
dependence for convenience. Again, this is exactly the solution found in
$\S$\ref{sec:wnlt}; similarly the solution at $\mathcal{O}(\epsilon^{5/2})$ is
as described in~$\S$\ref{sec:wnlt}. 

At $\mathcal{O}(\epsilon^{3})$, this analysis begins to differ from the one
in~$\S$\ref{sec:wnlt}. In particular, whereas in~$\S$\ref{sec:wnlt} the time
derivatives first appeared at $\mathcal{O}(\epsilon^{3})$, here we have no time
derivatives. Instead we obtain a fourth-order ODE for~$A_{2}$, namely
 \[
 16 A_{2}^{\prime\prime\prime\prime} = \left( 3 n_{3} + \tfrac{326}{81} n_{2}^{2} \right) |A_{2}|^{2} A_{2},
 \]
where the prime denotes differentiation with respect to~$X$. We obtain similar equations for the $A_{j}$, $j=3,\ldots,8$ at $\mathcal{O}(\epsilon^{2+j/2})$. For example, we find
 \[
 16 A_{3}^{\prime\prime\prime\prime} = \left( 3 n_{3} + \tfrac{326}{81} n_{2}^{2} \right) \left( 2 |A_{2}|^{2} A_{3} + A_{2}^{2} \bar{A}_{3} \right) + \frac{64}{243} i n_{2}^{2} |A_{2}|^{2} A_{2}^{\prime} + 32 i A_{2}^{\prime\prime\prime\prime\prime}
 \]
at $\mathcal{O}(\epsilon^{7/2})$. Continuing this process up to $\mathcal{O}(\epsilon^{6})$, we obtain all the required equations. We may then reconstitute these into one equation by defining
 \[
 Z(X) = \sum_{n=2}^{8} \epsilon^{(n-2)/2} A_{n}(X).
 \]
The resulting equation has derivatives up to eighth order. However, derivatives 
higher than fourth order appear at higher order than the $A''''$ term, and 
these can be eliminated by repeatedly differentiating the resulting equation 
and substituting back in to the equation. 
The resulting equation is then
 \begin{align}
 \label{eqn7_shs_z3}
 16 Z^{\prime\prime\prime\prime} &= \theta_{1} |Z|^{2} Z + \epsilon^{1/2} \left( \left( \theta_{2} + 4 i \theta_{1} \right) |Z|^{2} Z^{\prime} + 2 i \theta_{1} Z^{2} \bar{Z}^{\prime} \right) + \epsilon \Big( -\frac{5}{2} \theta_{1} Z^{2} \bar{Z}^{\prime\prime} \\ \nonumber
 &\qquad {} + \left( 2 i \theta_{3} - 5 \theta_{1} + \theta_{3} \right) \left( \bar{Z} (Z^{\prime})^{2} + |Z|^{2} Z^{\prime\prime} \right) + 2 \left( i \theta_{2} - 5 \theta_{1} \right) Z Z^{\prime} \bar{Z}^{\prime} \Big) + \mathcal{O}(\epsilon^{3/2}),
 \end{align}
where the $\theta_{i}$, $i=1,\ldots,3$ coefficients are
 \begin{equation*}
 \theta_{1} = 3 n_{3} + \frac{326}{81} n_{2}^{2},
 \qquad
 \theta_{2} = \frac{64}{243} i n_{2}^{2},
 \qquad
 \theta_{3} = - \frac{592}{729} n_{2}^{2}.
 \end{equation*}

To match to the scaled normal form equation (derived in the next section), we
require one more transformation. This is
 \begin{align}
 \label{eqn7_shs_rho}
 Z &= A + \epsilon^{2} \rho_{1} |A|^{2} A + \epsilon^{5/2} \left( \rho_{2} |A|^{2} A^{\prime} + \rho_{3} A^{2} \bar{A}^{\prime} \right) \\ \nonumber
 &\qquad {} + \epsilon^{3} \left( \rho_{4} |A|^{2} A^{\prime\prime} + \rho_{5} A^{2} \bar{A}^{\prime\prime} + \rho_{6} \bar{A} (A^{\prime})^{2} + \rho_{7} A A^{\prime} \bar{A}^{\prime} \right),
 \end{align}
where the $\rho_{i}$, $i=1,\ldots,6$ coefficients are to be determined through
the matching procedure. Under this transformation, \eqref{eqn7_shs_z3} becomes
 \begin{align} \label{eqn7_shs_a1}
 16 A^{\prime\prime\prime\prime} &= \theta_{1} |A|^{2} A + \epsilon^{1/2} \left( \left( \theta_{2} + 4 i \theta_{1} \right) |A|^{2} A^{\prime} + 2 i \theta_{1} A^{2} \bar{A}^{\prime} \right) + \epsilon \Big( -\frac{5}{2} \theta_{1} A^{2} \bar{A}^{\prime\prime} \\ \nonumber
 &\qquad + \left( 2 i \theta_{3} - 5 \theta_{1} + \theta_{3} \right) \left( \bar{A} (A^{\prime})^{2} + |A|^{2} A^{\prime\prime} \right) + 2 \left( i \theta_{2} - 5 \theta_{1} \right) A A^{\prime} \bar{A}^{\prime} \Big) + \mathcal{O}(\epsilon^{3/2}).
 \end{align}
We have continued up to $\mathcal{O}(\epsilon^{3})$, but the number of terms
in~\eqref{eqn7_shs_a1} quickly escalates to such an extent that it is not
instructive to include them all here; the truncated form of~\eqref{eqn7_shs_a1}
is sufficient for the argument presented below.

\subsubsection{Normal form scaling}
\label{sec:app:nf_asymptotics}

To match the scaling in~$\S$\ref{sec:app:model_asymptotics}, we write
 \[
(A(X),B(X),C(X),D(X)) = (\epsilon \tilde{A}(X),\epsilon^{3/2} \tilde{B}(X),\epsilon^{2} \tilde{C}(X),\epsilon^{5/2} \tilde{D}(X)) e^{ix}, 
 \]
where, as before, $X = \sqrt{\epsilon} x$, and we have factored out an $e^{ix}$
dependence. With this, the normal form for
$A_{x}$,~\eqref{eq:app:normal_form_version_2a}, becomes
 \[
 A_{x} = \epsilon (i \tilde{A} + \sqrt{\epsilon} \tilde{A}_{X}) e^{ix} = \left( i \epsilon \tilde{A} + \epsilon^{3/2} \tilde{B} + i \epsilon \tilde{A} \epsilon P ( \epsilon^{2} \tilde{c_{1}},\epsilon^{5/2} \tilde{c_{3}},\epsilon^{3} \tilde{c_{5}},\epsilon^{7/2} \tilde{c_{7}},\epsilon \tilde{w_{1}},\epsilon^{3/2} \tilde{w_{2}} ) \right) e^{ix}.
 \]
Cancelling the common factor $e^{ix}$, subtracting $i \epsilon \tilde{A}$ from
both sides and dropping the tildes, we have
 \begin{align} \nonumber
 \epsilon^{3/2} A_{X} 
 &= \epsilon^{3/2} B + i \epsilon^{3} P_{1} |A|^{2} A - \epsilon^{7/2} P_{2} A \left( A \bar{B} - \bar{A} B \right) + i \epsilon^{4} A \left( A \bar{C} - B \bar{B} + \bar{A} C \right) \\ \nonumber
 &\qquad {} - \epsilon^{9/2} P_{4} A \left( A \bar{D} - B \bar{C} + \bar{B} C - \bar{A} D \right)  - \epsilon^{4} T_{2} \bar{A} \left( B^{2} - 2 A C \right) \\ \label{eqn7_ax}
 &\qquad \qquad {} + 3 \epsilon^{9/2} T_{1} \left( \bar{A} \left(3 A D - B C \right) + \bar{B} \left( B^{2} - 2 A C \right) \right).
 \end{align}
We may cast this in appropriate form for the analysis by dividing through by a factor of $\epsilon^{3/2}$ and rearranging, such that we have
\begin{equation} \label{eqn7_nfs_b1}
 B = A_{X} + \mathcal{O}(\epsilon^{3/2}).
\end{equation}
We do not explicitly write out all the terms for the sake of clarity in explaining the procedure.

Similarly, from~\eqref{eq:app:normal_form_version_2} we have
 \begin{equation} \label{eqn7_nfs_c1}
 C = B_{X} - \epsilon Q_{1} |A|^{2} A + \mathcal{O}(\epsilon^{3/2})
 \end{equation}
and
 \begin{equation}  \label{eqn7_nfs_d1}
 D = C_{X} - i \epsilon^{1/2} R_{1} |A|^{2} A + \epsilon \left( R_{2} A^{2} \bar{B} - \left( Q_{1} + R_{2} \right) |A|^{2} B \right) + \mathcal{O}(\epsilon^{3/2}). 
 \end{equation}
We also obtain an equation for $D_{X}$, which we leave in the form
 \begin{align} \label{eqn7_nfs_d1X}
 D_{X} &= S_{1} |A|^{2} A + i \epsilon^{1/2} \left( S_{2} A^{2} \bar{B} + \left( R_{1} - S_{2} \right) |A|^{2} B \right) \\ \nonumber
 &\qquad {} + \epsilon \left( S_{3} A^{2} \bar{C} - \left( R_{2} + S_{3} \right) A B \bar{B} + \left( R_{2} + T_{8} \right) \bar{A} B^{2} \right. \\ \nonumber
 &\qquad \qquad \left. {}+ \left( S_{3} + Q_{1} - 2 T_{8} \right) |A|^{2} C \right) + \mathcal{O}(\epsilon^{3/2}).
 \end{align}

The next step is to differentiate~\eqref{eqn7_nfs_b1} with respect to~$X$, giving
\begin{equation} \label{eqn7_nfs_b2}
 B_{X} = A_{XX}  + \mathcal{O}(\epsilon^{3/2}).
\end{equation}
Substituting~\eqref{eqn7_nfs_b2} into~\eqref{eqn7_nfs_c1}, we have
 \begin{equation} \label{eqn7_nfs_c2}
 C = A_{XX}  - \epsilon Q_{1} |A|^{2} A + \mathcal{O}(\epsilon^{3/2}).
 \end{equation}
This procedure of differentiation and substitution is repeated twice more. Differentiating~\eqref{eqn7_nfs_c2} with respect to~$X$ we have
\begin{equation} \label{eqn7_nfs_c3}
 C_{X} = A_{XXX} - \epsilon Q_{1} \left( 2 |A|^{2} A_{X} + A^{2} \bar{A}_{X} \right) + \mathcal{O}(\epsilon^{3/2}).
\end{equation}
Before substituting~\eqref{eqn7_nfs_c3} into~\eqref{eqn7_nfs_d1}, we first
replace the $B$ and $\bar{B}$ terms
using~\eqref{eqn7_nfs_b1}. This yields
 \begin{equation} \label{eqn7_nfs_d2}
 D = C_{X} - i \epsilon^{1/2} R_{1} |A|^{2} A + \epsilon \left( R_{2} A^{2} \bar{A}_{X} - \left( Q_{1} + R_{2} \right) |A|^{2} A_{X} \right) + \mathcal{O}(\epsilon^{3/2}).
 \end{equation}
Now substituting~\eqref{eqn7_nfs_c3} into~\eqref{eqn7_nfs_d2} we have
\begin{align} \label{eqn7_nfs_d3}
 D &= A_{XXX} - \epsilon Q_{1} \left( 2 |A|^{2} A_{X} + A^{2} \bar{A}_{X} \right) - i \epsilon^{1/2} R_{1} |A|^{2} A \\ \nonumber
 &\qquad + \epsilon \left( R_{2} A^{2} \bar{A}_{X} - \left( Q_{1} + R_{2} \right) |A|^{2} A_{X} \right) + \mathcal{O}(\epsilon^{3/2}) \\ \nonumber
 &= A_{XXX} - i \epsilon^{1/2} R_{1} |A|^{2} A + \epsilon \left( \left( R_{2} - Q_{1} \right) |A|^{2} A_{X} - \left( R_{2} + 3 Q_{1} \right) A^{2} \bar{A}_{X} \right) + \mathcal{O}(\epsilon^{3/2}).
\end{align}
Differentiating one final time, we have
\begin{align} \label{eqn7_nfs_d4}
 D_{X} &= A_{XXXX} - i \epsilon^{1/2} R_{1} \left( 2 |A|^{2} A_{X} + A^{2} \bar{A}_{X} \right) + \epsilon \left( \left( R_{2} - 5 Q_{1} \right) A A_{X} \bar{A}_{X} \right. \\ \nonumber
 &\qquad \left. + \left( R_{2} - Q_{1} \right) A^{2} \bar{A}_{XX} - \left( R_{1} + 3 Q_{1} \right) \left( \bar{A} A_{X}^{2} + |A|^{2} A_{XX} \right) \right)  + \mathcal{O}(\epsilon^{3/2}).
\end{align}

We now notice that we have two equations for $D_{X}$, \eqref{eqn7_nfs_d1X}
and~\eqref{eqn7_nfs_d4}, and thus equating the two will yield an equation 
for~$A_{XXXX}$. Before we do this, we first should use~\eqref{eqn7_nfs_b2}
and~\eqref{eqn7_nfs_c2} to replace the $B$, $\bar{B}$, $C$ and $\bar{C}$ terms
in~\eqref{eqn7_nfs_d1X}. This gives
 \begin{align} \label{eqn7_nfs_d5}
 D_{X} &= S_{1} |A|^{2} A + i \epsilon^{1/2} \left( S_{2} A^{2} \bar{A}_{X} + \left( R_{1} - S_{2} \right) |A|^{2} A_{X} \right) \\ \nonumber
 &\qquad + \epsilon \left( S_{3} A^{2} \bar{A}_{XX} - \left( R_{2} + S_{3} \right) A A_{X} \bar{A}_{X} + \left( R_{2} + T_{8} \right) \bar{A} A_{X}^{2} \right. \\ \nonumber
 &\qquad \qquad \left. + \left( S_{3} + Q_{1} - 2 T_{8} \right) |A|^{2} A_{XX} \right) + \mathcal{O}(\epsilon^{3/2}).
\end{align}

Equating~\eqref{eqn7_nfs_d4} and~\eqref{eqn7_nfs_d5}, we find
\begin{align} \label{eqn7_nfs_eqn}
 A_{XXXX} &= S_{1} |A|^{2} A + \epsilon^{1/2} i ((3 R_{1} - S_{2}) |A|^{2} A_{X} + (S_{2} + R_{1}) A^{2} \bar{A}_{X} ) \\ \nonumber
           &\qquad + \epsilon ((2R_{2} + T_{8} + 3 Q_{1})\bar{A} A_{X}^{2} + (5 Q_{1} - 2 R_{2} - S_{3}) A A_{X} \bar{A}_{X} \\ \nonumber
            &\qquad + (4 Q_{1} + S_{3} + R_{2} - 2 T_{8}) |A|^{2} A_{XX} + (S_{3} + Q_{1} - R_{2}) A^{2} \bar{A}_{XX}) + \mathcal{O}(\epsilon^{3/2}).
\end{align}

As in~$\S$\ref{sec:app:model_asymptotics}, continuing up to
$\mathcal{O}(\epsilon^{3})$ yields a very large number of terms, so we truncate
at the same order as~\eqref{eqn7_shs_a1}. This is sufficient for the
coefficient matching in the next section.

\subsubsection{Order-by-order matching}
\label{sec:app:order_by_order}

We now match the two equations~\eqref{eqn7_shs_a1} and~\eqref{eqn7_nfs_eqn} at
each power of~$\epsilon^{1/2}$ to recover the normal form
coefficients. The subscript $X$ in~\eqref{eqn7_nfs_eqn} and superscript prime
in~\eqref{eqn7_shs_a1} are now equivalent. At leading order the solution is
immediately recovered, and we have
 \[
 S_{1} = \frac{3}{16} n_{3} + \frac{163}{648} n_{2}^{2}. 
 \]
This is the same as calculated via the nonlinear coordinate transform method,
as given in~\eqref{eqn7_nf_coeffs1}.

Proceeding to $\mathcal{O}(\epsilon^{1/2})$ we have to solve the equations
 \[
 2 i \theta_{1} = i \left( R_{1} + S_{2} \right),
 \qquad
 \theta_{2} + 4 i \theta_{1} = i \left( 3 R_{1} + S_{2} \right). 
 \]
The solution to these two equations is
 \[
 R_{1} = \frac{9}{32} n_{3} + \frac{1483}{3888} n_{2}^{2},
 \qquad
 S_{2} = \frac{3}{32} n_{3} + \frac{473}{3888} n_{2}^{2},
 \]
again, the same as~\eqref{eqn7_nf_coeffs1}.

At the next order, $\mathcal{O}(\epsilon)$, we have the four equations
\begin{align} \nonumber
 -\frac{5}{2} \theta_{1} &= S_{3} + Q_{1} - R_{2}, \\ \nonumber
2 i \theta_{2} - 5 \theta_{1} + \theta_{3} &= 4 Q_{1} + S_{3} + R_{2} - 2 T_{8}, \\ \nonumber
2 i \theta_{2} - 5 \theta_{1} + \theta_{3} &= 2 R_{2} + T_{8} + 3 Q_{1}, \\ \nonumber
2 \left( i \theta_{2} - 5 \theta_{1} \right) &= 5 Q_{1} - 2 R_{2} - S_{3}.
\end{align}
Solving these four equations simultaneously, we find
\begin{align} \nonumber
  Q_{1} = -\frac{75}{224} n_{3} - \frac{37907}{81648} n_{2}^{2}&, \qquad R_{2} = \frac{25}{224} n_{3} + \frac{10657}{81648} n_{2}^{2}, \\ \nonumber
  S_{3} = -\frac{5}{224} n_{3} - \frac{103}{3024} n_{2}^{2} &, \qquad T_{8} = -\frac{5}{32} n_{3} - \frac{815}{3888} n_{2}^{2}.
\end{align}
Comparing with~\eqref{eqn7_nf_coeffs1}, we see that there is a disagreement in
three out of the four terms. $T_{8}$~agrees, whereas the other three agree only
in the dependence on the cubic nonlinearity coefficient~$n_{3}$.

The differences in $Q_{1}$, $R_{2}$ and $S_{3}$ are $2 n_{2}^{2}/7$, $-3
n_{2}^{2}/7$, $2 n_{2}^{2}/7$ respectively. This trend continues at higher
orders, \textit{i.e.}, the normal form coefficients found by the asymptotic
scaling method agree with those found by the nonlinear coordinate transform
method only in the dependence on $n_{3}$. We note that any extra terms included
in the transformation~\eqref{eqn7_shs_rho} to try and rectify this would depend
on $\theta_{1}$, and thus change the $n_{3}$-dependence of the normal form
coefficients.

A possible explanation for this discrepancy is the non-uniqueness of the normal
form. In the analysis of the Swift--Hohenberg equation, the equivalent linear
transformation to~\eqref{eq:app:linear_transform} is more generally a
two-parameter family of transformations \citep{Burke2007a}. Fixing the value of
these parameters at linear order determines the values of the normal form
coefficients at higher order. By extension, we similarly will have a
four-parameter family of transformations. Whereas in the Hamiltonian--Hopf case
setting these parameters to determine the linear coordinate transform
automatically determines the normal form coefficients at higher order, that is
not necessarily the case here. There may be extra parameters that have
implicitly been set differently by the two different methods.

Similarly, it is possible the $w_{1}$ and $w_{2}$ terms could have introduced an extra hidden parameter into the system, and the value of this parameter chosen by the two methods is not consistent.

\subsection{First integrals of the normal form}

The normal form for the Hamiltonian--Hopf bifurcation has two integrals which
allow one to determine geometrically the solutions of the Swift--Hohenberg
equation \citep{Iooss1993,Iooss1998a,Woods1999}. So, for the normal form in our 
case to be of any use, we need to find integrals
of the normal form~\eqref{eq:app:normal_form_version_2}.

One of the integrals of the normal form for the Hamiltonian--Hopf bifurcation
is also an integral of the characteristic system used to derive the normal
form. Why this should be the case is not obvious. Nonetheless, we might hope
that one of the integrals (or some combination of the integrals) in
Table~\ref{tab:integrals} is also be an integral for the normal
form~\eqref{eq:app:normal_form_version_2}. 

We first concentrate on the linear
terms of the normal form~\eqref{eq:app:nf_linear}, and consider the integral
$c_{7} = i \left( A \bar{D} - B \bar{C} + \bar{B} C - \bar{A} D \right)$.
Differentiating $c_{7}$  with respect to~$x$, we have
 \begin{align*} 
 \frac{d c_{7}}{d x} &= \frac{d}{dx} \left( A \bar{D} - B \bar{C} + \bar{B} C - \bar{A} D \right) \\ 
 &= A \frac{d \bar{D}}{d x} + \bar{D} \frac{d A}{d x} - B \frac{d \bar{C}}{d x} - \bar{C} \frac{d B}{d x} + C \frac{d \bar{B}}{d x} + \bar{B} \frac{d C}{d x} - \bar{A} \frac{d D}{d x} - D \frac{d \bar{A}}{d x} \\ 
 &= A \left(- i \bar{D} \right) + \bar{D} \left( i A + B \right) - B \left( - i \bar{C} + \bar{D} \right) - \bar{C} \left( i B + C \right) + C \left( -i \bar{B} + \bar{C} \right) \\ 
 &\qquad + \bar{B} \left( i C + D \right) - \bar{A} \left( i D \right) - D \left( -i \bar{A} + \bar{B} \right) \\ 
 &= 0,
 \end{align*}
using~\eqref{eq:app:nf_linear}. Thus $c_{7}$ is an integral
of the linear normal form~\eqref{eq:app:nf_linear}. There are three similar 
integrals of the linear normal form, to wit
 \begin{align*} 
 \frac{d}{dx} \left( B \bar{D} - C \bar{C} + \bar{B} D \right) &= B \left( - i \bar{D} \right) + \bar{D} \left( i B + C \right) - C \left( - i \bar{C} + \bar{D} \right) - \bar{C} \left( i C + D \right)\\ 
 &\qquad + \bar{B} \left( i D \right) + D \left( - i \bar{B} + \bar{C} \right) = 0,
\end{align*}
\begin{equation*} 
 \frac{d}{dx} \left( C \bar{D} - \bar{C} D \right) = C \left( - i \bar{D} \right) + \bar{D} \left( i C + D \right) - \bar{C} \left( i D \right) - D \left( -i \bar{C} + \bar{D} \right) = 0,
\end{equation*}
and
\begin{equation*} 
 \frac{d}{dx} |D|^{2} = D \left(-i \bar{D} \right) + \bar{D} \left( i D \right) = 0.
\end{equation*}
However, none of these four integrals of the linear normal form are integrals
of the nonlinear normal form, nor is any combination of them. This failure to
extend to the nonlinear normal form is entirely down to the presence of the
$w_{1}$ and $w_{2}$ terms in the normal form. In particular, it is the fact
that $w_{1}$ and~$w_{2}$ are not real that the extension fails. A wider search
for other possible forms for integrals did not uncover any.

The absence of integrals of the normal form prevents the extension of
the geometric analysis of the Hamiltonian--Hopf bifurcation to our degenerate
situation. A possible explanation for the lack of integrals of the normal form
is the non-uniqueness of the normal form: other choices of which terms
to  keep in the polynomial~$\mathbf{P}$ in~\eqref{eq:app:ending_point} might
lead to an integrable normal form. It would be interesting to pursue this 
further.


\bibliographystyle{IMANUM-BIB} 
\bibliography{allrefs} 

\end{document}